\documentclass[%
 reprint,
 amsmath,amssymb,
 aps,
]{revtex4-2}

\usepackage{float}
\makeatletter
\let\newfloat\newfloat@ltx
\makeatother

\usepackage{graphicx}
\usepackage{dcolumn}
\usepackage{bm}
\usepackage{graphicx}
\usepackage{amsmath}
\usepackage{amssymb}
\usepackage{subfig}
\usepackage{subcaption}
\usepackage{qcircuit}
\usepackage{algorithm}
\usepackage{algpseudocode}
\usepackage{cprotect}
\usepackage{lipsum}
\usepackage{booktabs}
\usepackage{array}
\usepackage{comment}
\usepackage{amsmath, amsthm, amssymb}
\usepackage{braket}
\usepackage{bm}
\usepackage{physics}
\usepackage{graphics}
\usepackage[dvipsnames]{xcolor}

\usepackage{caption}

\theoremstyle{definition}

\newcommand{\tet}{\bm{\theta}}
\newcommand{\Break}{\State \textbf{break} }

\usepackage[makeroom]{cancel}

\makeatletter
\@ifundefined{sf@counterlist}{\newcommand{\sf@counterlist}{}}{}
\makeatother
\usepackage[colorlinks=true, allcolors=blue]{hyperref}

\usepackage{cleveref}

\begin{document}

\title{Enhancing Hybrid Methods in Parameterized Quantum Circuit Optimization}

\author{Joona V. Pankkonen}
\email{Corresponding author: joona.pankkonen@aalto.fi}
\author{Matti Raasakka}
\author{Andrea Marchesin}
\author{Ilkka Tittonen}
 \affiliation{%
 Micro and Quantum Systems group, Department of Electronics and Nanoengineering,\\Aalto University, Finland
}%

\date{\today}

\begin{abstract}
Parameterized quantum circuits (PQCs) play an essential role in the application of variational quantum algorithms (VQAs) in noisy intermediate-scale quantum (NISQ) devices. The PQCs are a leading candidate to achieve a quantum advantage in NISQ devices and have already been applied in various domains such as quantum chemistry, quantum machine learning, combinatorial optimization, and many others. There is no single definitive way to optimize PQCs. The most commonly used methods are based on computing the gradient via the parameter-shift rule to use classical gradient descent (GD) optimizers like Adam, stochastic GD, and others.  In addition, sequential single-qubit optimizers have been proposed, such as \verb|Rotosolve|, Free-Axis Selection (\verb|Fraxis|), Free-Quaternion Selection (\verb|FQS|), and hybrid algorithms from the aforementioned optimizers. We further develop hybrid algorithms than those represented in the previous work by drawing inspiration from the early stopping method used in classical machine learning. The switch between the optimizers depends on the previous cost function values compared to the previous ones. We introduce two new hybrid algorithms that are more robust and scalable, and they outperform previous hybrid methods in terms of convergence towards the global minima across various cost functions. In addition, we find that they are feasible for NISQ devices with different noise profiles.
\end{abstract}

\maketitle


\section{Introduction}
Parameterized quantum circuits (PQCs) play an essential role in the application of variational quantum algorithms (VQAs)~\cite{sim2019expressibility} in noisy intermediate-scale quantum (NISQ) devices~\cite{preskill_nisq}. The PQCs are a leading candidate to achieve a quantum advantage in NISQ devices~\cite{cerezo2021variational, peruzzo2014variational} and have already been applied in various domains such as quantum chemistry~\cite{bauer2020quantum, singh2023benchmarking, delgado2021variational, motta2022emerging, cao2019quantum}, quantum machine learning~\cite{liu2021hybrid, arthur2022hybrid, jerbi2023quantum, shi2022parameterized}, combinatorial optimization~\cite{comb_opt_cite1, comb_opt_cite2, comb_opt_cite3}, and many others~\cite{li2022concentration, sauvage2024building, haug2021capacity, nakaji2022approximate, xiao2024quantum}. PQCs consist of quantum gates that have tunable parameters. By using a classical feedback loop, we can optimize the parameters using a classical computer by evaluating a cost function by measuring the PQC.

There is no single definitive way to optimize PQCs. The most commonly used methods are based on computing gradient via parameter-shift rule~\cite{crooks2019gradients, wierichs2022general, banchi2021measuring} to use classical gradient descent (GD) optimizers like Adam~\cite{kingma2014adam}, stochastic GD, and others~\cite{qian1999momentum, nesterov1983method, zeiler2012adadelta, duchi2011adaptive}. The Quantum Natural Gradient (QNG)~\cite{stokes2020quantum} and its variations~\cite{koczor2022quantum, halla2025quantum, halla2025modified, qi2024federated, sasaki2024quantum} have further improved the classical GD methods by incorporating Quantum Information Geometry into the optimization process, surpassing the classical GD and Adam optimizers in performance. In addition, gradient-free sequential single-qubit optimizers have been proposed, such as \verb|Rotosolve|~\cite{Ostaszewski_2021}, Free-Axis Selection (\verb|Fraxis|)~\cite{fraxis}, Free-Quaternion Selection (\verb|FQS|)~\cite{fqs}, as well as random axis initialization method and hybrid algorithms from the aforementioned optimizers~\cite{pankkonen2025improvingvariationalquantumcircuit}.

We develop hybrid algorithms further than the ones represented in~\cite{pankkonen2025improvingvariationalquantumcircuit}, where the combination of optimizers \verb|Rotosolve| and \verb|FQS| was proposed to enhance the optimization of the circuit. Previous methods were based on probabilistic or iteration-specific ways to optimize the gates in the PQC. The probabilistic hybrid optimized each gate with a less expressive optimizer with the probability $p$, and otherwise it was optimized with the more expressive optimizer. We define expressivity as how well the optimizer is able to represent complex states in the Hilbert space. The iteration-specific hybrid, as the name suggests, was based on the iterations in PQC optimization, where one iteration means optimizing all gates sequentially exactly once. The iteration hybrid used the more expressive optimizer for every $N$-th iteration, and otherwise it used the less expressive optimizer. In this work, we create more robust and scalable hybrid algorithms based on the previous values of the cost function and draw inspiration and connection to early stopping methods~\cite{early_stopping, early_stopping_NN, mahsereci2017early, li2021self} used in classical machine learning (ML). Early stopping has been shown to prevent overfitting of ML models~\cite{rice2020overfitting, early_stopping_NN}. As in classical ML, when the model training is stopped as the validation accuracy plateaus or starts to get worse, we switch from a less expressive optimizer to a more expressive one. In this work, we implement a method that uses cost function values as an analogous measure of training accuracy and use the idea of early stopping patience~\cite{patience, paguada2023being} to implement the point where we switch optimizers used in PQC optimization. In this work, patience is defined as the number of times we allow the cost function to vary within a given threshold. Additionally, in our second method, we examine the cost function average within a given interval and switch optimizers if the absolute difference between the newest cost function value and the cost function average is below a certain threshold. We demonstrate efficiency, scalability, and robustness across various system sizes and cost functions through numerical experiments. The hybrid methods proposed in this work show better and faster convergence in terms of circuit evaluations for the Heisenberg and Fermi-Hubbard models across different system sizes when compared to the probabilistic and iteration-specific hybrids proposed in previous work. 

This work is structured as follows. First, in Sec.~\ref{Methods_section}, we go through the optimization of PQC in general. Then, we present two new hybrid methods that utilize cost function values to determine which optimizer to use. In Sec.~\ref{results_section}, we provide numerical experiments for the one-dimensional Heisenberg and Fermi-Hubbard model. In addition, we examined the scalability of the Heisenberg model with various system sizes. Finally, we tested fidelity maximization for 4-qubit random quantum states. In Sec.~\ref{conclusion_section}, we conclude our work and discuss possible extensions to this work.

\section{Methods} \label{Methods_section}

\subsection{Optimizing Parameterized Quantum Circuits}

We start by examining a PQC, which is represented by a unitary $U(\tet)$ consisting of real-valued parameters $\tet$ \cite{cerezo2021variational}. Typically, a PQC is constructed by applying a layer of parameterized single-qubit gates followed by an entangling layer of two-qubit gates. Commonly controlled-Z or CNOT gates are used in the entangling layer. This pattern of parameterized and entangling gates is repeated $L$ times, and the PQC can be expressed as 
\begin{equation}\label{unitaries_eq}
    U(\tet) = U_{L-1}(\tet_{L-1})\cdots U_1(\tet_1)U_0(\tet_0),
\end{equation}
where $U_l(\tet_l)$ corresponds to a $l$-th unitary in the circuit. Here we start the index from zero for the unitary $U(\tet)$ for cleaner notation.

\begin{figure}
    \[
    \Qcircuit @C=2em @R=.9em {
    & \mbox{$L$ layers} & & &  \\
    & & & & & &  \\
     \lstick{\ket{0}_1} & \gate{U_{ln + 1}\Bigl(\theta_{ln + 1}\Bigr)}  &  \ctrl{0} & \qw & \qw &  \meter\\
     \lstick{\ket{0}_2}&  \gate{U_{ln + 2}\Bigl(\theta_{ln + 2}\Bigr)} &  \ctrl{-1} & \ctrl{1} & \qw  &   \meter\\
     \lstick{\ket{0}_3}&  \gate{U_{ln + 3}\Bigl(\theta_{ln + 3}\Bigr)}  &  \ctrl{0} & \ctrl{0} & \qw &   \meter\\
     \lstick{\ket{0}_4}&  \gate{U_{ln + 4}\Bigl(\theta_{ln + 4}\Bigr)}  &  \ctrl{-1} & \ctrl{0} & \qw  &  \meter\\
     \lstick{\ket{0}_5}&  \gate{U_{ln + 5}\Bigl(\theta_{ln + 5}\Bigr)} &  \qw & \ctrl{-1} & \qw &  \meter  \gategroup{3}{2}{7}{4}{1.2em}{--} 
    }
    \]
    \caption{Ansatz circuit design for PQC optimization.}
    \label{Ansatz_circuit_image}
\end{figure}
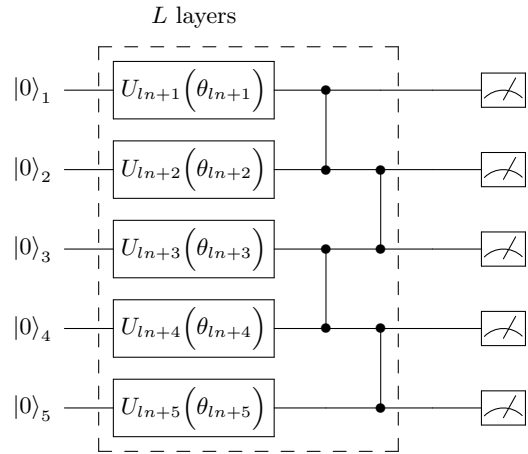

For a $n$-qubit system, the $l$-th unitary can be expressed as follows
\begin{equation}\label{unitary_layer}
    U_l(\tet_l) = W_l \left( \bigotimes_{k=1}^n e^{-i \theta_{ln + k} H_{ln + k} / 2} \right),
\end{equation}\\[0.05cm]
where $k$ indexes the individual qubits. $H_{ln + k}$ is a Hermitian operator that acts on the $k$-th qubit in the $l$-th layer, and $\theta_{ln + k}$ is the parameter of the single-qubit gate that acts on the $k$-th qubit in the $l$-th layer. The $W_l$ in Eq.~(\ref{unitaries_eq}) denotes the entangling layer consisting of controlled-Z or CNOT gates. An illustration of this kind of PQC is shown in Fig.~\ref{Ansatz_circuit_image}, emphasizing the structure of the $l$-th unitary $U_l(\bm{\theta}_l)$.

To optimize the PQC, we first need to define a cost function $C(\tet)$ that depends on the parameters of the single-qubit gate $\tet$ \cite{cost_function}. The cost function provides us with a quantitative measure of how well the PQC performs on the given task. The cost function used in this work is the expectation value of a Hermitian observable $\hat{M}$, which is written as follows
\begin{equation}
    \langle \hat{M} \rangle = \Tr\left(\hat{M} U(\tet) \rho_0 U(\tet)^\dagger \right),
\end{equation}\\
where $\rho_0$ denotes the initial state of the PQC. In this work, we set $\rho_0 = \ket{\bm{0}} \bra{\bm{0}}$, where $\ket{\bm{0}}$ is the $n$-qubit state $\ket{0}^{\otimes n}$.

The ansatz circuit consisting of a total of $L$ layers has $Ln$ parameters $\tet = (\theta_1,\ldots,\theta_{Ln})$. The $d$-th single-qubit gate $U_d$ $(d = 1,\ldots ,Ln)$ is be expressed as 
\begin{equation}\label{single_qubit_gate}
    U_d(\theta_d) = \cos \left( \frac{\theta_d}{2} \right) I - i \sin \left( \frac{\theta_d}{2} \right) H_d,
\end{equation}
where $I$ is the $2\times 2$ identity matrix. $H_d$ is the Hermitian unitary generator that satisfies $H_d^2 = I$. To operate $U_d$ in the PQC on the $k$-th qubit, we need to express it as a $2^n \times 2^n$ unitary matrix since the PQC is a $2^n \times 2^n$ unitary matrix. We achieve this by applying a tensor product of $2\times 2$ identity matrices to $U_d$. To be precise, before the gate $U_d$ precedes a tensor product of $k-1$ identity matrices and is followed by a tensor product of $n-k$ identity matrices. Then, we may express the newly obtained form of $U_d$ as 
\begin{eqnarray} \label{dth_gate}
    U_d '(\theta_d) = \ &&I \otimes  \cdots \otimes I \otimes \left[ \cos\left(\frac{\theta_d}{2} \right)I - i\sin\left(\frac{\theta_d}{2} \right)H_d \right] \nonumber \\ && \otimes I \otimes \cdots \otimes I.
\end{eqnarray}
Now by applying Eq.~(\ref{dth_gate}) to Eq.~(\ref{unitary_layer}) we can write the cost function as
\begin{widetext}
\begin{eqnarray}
    \langle \hat{M} \rangle = \Tr\left(\hat{M}W_{L-1} U_{Ln}(\theta_{Ln})\cdots W_l U_{(l-1)n}(\theta_{(l-1)n})\cdots U_1(\theta_1) \rho_0  U_1(\theta_1)^\dagger \cdots U_{(l-1)n}(\theta_{(l-1)n})^\dagger W_l \cdots U_{Ln}(\theta_{Ln})^\dagger W_{L-1} \right).  \qquad
\end{eqnarray}
\end{widetext}
Then, we define quantum circuits that come before and after the gate $U_d'$ as $V_1$ and $V_2$, respectively. After that, we can express the cost function in a more compact form as follows
\begin{equation}
     \langle \hat{M} \rangle = \Tr\left( \hat{M} V_1 U_d'(\theta_d) V_2 \rho_0 V_2^\dagger U_d'(\theta)^\dagger V_1^\dagger  \right).
\end{equation}
Furthermore, by utilizing the cyclic property of the trace operation and defining 
\begin{align}
    M &\equiv V_1^\dagger \hat{M} V_1, \\
    \rho &\equiv V_2 \rho_0 V_2^\dagger,
\end{align}
we obtain the commonly used expression for the cost function in PQC optimization~\cite{Ostaszewski_2021, fraxis, fqs}
\begin{equation}\label{cost_function_final}
    \langle M \rangle = \Tr\left(M U_d'(\theta_d)\rho U_d'(\theta)^\dagger \right).
\end{equation}
This form is used to fix all single-qubit gates in the PQC except the $d$-th one, which gradient-free sequential single-qubit gate optimizers \verb|Rotosolve|~\cite{Ostaszewski_2021}, \verb|Fraxis|~\cite{fraxis}, and \verb|FQS|~\cite{fqs} use in PQC optimization. Additionally, this is also used in hybrid methods in Ref.~\cite{pankkonen2025improvingvariationalquantumcircuit}, where the optimizers \verb|Rotosolve| and \verb|FQS| were combined into a hybrid algorithm that utilizes the fast convergence of \verb|Rotosolve| and the superior expressivity of \verb|FQS|. Next, we present two new hybrid methods for PQC optimization.

\subsection{Hybrid Algorithms}

Next, we introduce hybrid algorithms more robust and deterministic than those presented in previous work in~\cite{pankkonen2025improvingvariationalquantumcircuit}. In previous work, hybrid algorithms composed of \verb|Rotosolve| and \verb|FQS| were based on gate-wise optimization, where the individual gate is optimized with \verb|Rotosolve| with probability $p$ and otherwise with \verb|FQS|. Additionally, an iteration-specific hybrid algorithm was proposed, where every $N$-th iteration, the PQC was optimized by \verb|FQS|, and otherwise by \verb|Rotosolve|. In this work, we use \verb|Rotosolve| at the beginning of optimization and then switch to \verb|FQS| when the set criterion is met.

We draw inspiration from the early stopping methods~\cite{early_stopping, early_stopping_NN, mahsereci2017early, li2021self, patience, paguada2023being} used in classical ML, where model training is stopped to prevent overfitting. To be precise, we implement the early stopping method in hybrid algorithms based on how the value of the cost function changes compared to the previous iteration. If the cost function changes less than a given threshold $E_t$, we increment the patience counter by one. When the set patience counter reaches the given limit $P$, we change from a less expressive algorithm $\mathcal{A}$ to a more expressive one $\mathcal{B}$. We describe this method in Algorithm~\ref{algo1}.

\begin{algorithm}
\caption{Early Stopping for Hybrid Algorithms}\label{algo1}
\begin{algorithmic}[1]
\State \textbf{Inputs}: A Parameterized Quantum Circuit $U$ with fixed architecture, Hermitian measurement operator as the cost function, heuristically selected stopping criterion, less and more expressive optimizers $\mathcal{A}$ and $\mathcal{B}$ (e.g. \verb|Rotosolve| and \verb|FQS|, respectively).
\State Initialize a fixed threshold $E_t$.
\State Initialize a maximum value $P$ for patience counter.
\State Initialize patience counter to zero.
\State Initialize the parameters, $\theta_d \in (-\pi, \pi]$ for $d = 1, \ldots, Ln$ heuristically or at random for optimizer $\mathcal{A}$.

\For{$d = 1, \ldots, Ln$}
    \State Fix all gates except the $d$-th one.
    \State Optimize the $d$-th gate $U_d$ with optimizer $\mathcal{A}$.
    \State Compute $\Delta\expval{M} \leftarrow \abs{\expval{M}_{prev} - \expval{M}_{new}}$
    \If {$\Delta \expval{M} < E_t$ } 
    \State $patience \leftarrow patience + 1$
    \If {$patience$ = $P$ }
    \Break
    \EndIf
    \EndIf
\EndFor
\State Switch from less expressive optimizer $\mathcal{A}$ to more expressive optimizer $\mathcal{B}$.
\Repeat
    \For{$d = 1, \ldots, Ln$}
        \State Fix all gates except the $d$-th one.
        \State Optimize the $d$-th gate $U_d$ with optimizer $\mathcal{B}$.
    \EndFor
\Until{stopping criterion is met.}
\end{algorithmic}
\end{algorithm}

In addition, we propose another method to determine when to switch from the least expressive optimizer $\mathcal{A}$ to a more expressive optimizer $\mathcal{B}$. Instead of determining the switch of optimizers by patience and threshold, we focus on computing the average from the previous cost function values and comparing it to the newly computed cost function value. That is, for a fixed window of length $w$, we compute the average cost function value $\expval{M}_{avg}$ and compute the absolute difference to the newly computed cost function value $\expval{M}$. If this difference exceeds the given threshold $E_t$, then we switch from the optimizer $\mathcal{A}$ to the optimizer $\mathcal{B}$ for the rest of the optimization process. The algorithm is described in detail in Algorithm~\ref{algo2}.

\begin{algorithm}
\caption{Convergence Comparison with Cost Function Average}\label{algo2}
\begin{algorithmic}[1]
\State \textbf{Inputs}: A Parameterized Quantum Circuit $U$ with fixed architecture, Hermitian measurement operator as the cost function, heuristically selected stopping criterion, less and more expressive optimizers $\mathcal{A}$ and $\mathcal{B}$ (e.g. \verb|Rotosolve| and \verb|FQS|, respectively).
\State Initialize a fixed threshold $E_t$ for cost function difference.
\State Initialize a fixed window $w$ to determine the number of previous cost function values where the average is computed.
\State Initialize the parameters $\theta_d \in (-\pi, \pi]$ for $d = 1, \ldots, Ln$ heuristically or at random for optimizer $\mathcal{A}$.
\For{$d = 1, \ldots, Ln$}
    \State Fix all gates except the $d$-th one.
    \State Optimize the $d$-th gate $U_d$ with optimizer $\mathcal{A}$.
    \State Compute $\Delta\expval{M} \leftarrow \abs{\expval{M}_{avg} - \expval{M}}$
    \If {$\Delta \expval{M} < E_t$ } 
    \Break
    \EndIf
\EndFor
\State Switch from less expressive optimizer $\mathcal{A}$ to more expressive optimizer $\mathcal{B}$.
\Repeat
    \For{$d = 1, \ldots, Ln$}
        \State Fix all gates except the $d$-th one.
        \State Optimize the $d$-th gate $U_d$ with optimizer $\mathcal{B}$.
    \EndFor
\Until{stopping criterion is met.}
\end{algorithmic}
\end{algorithm}

\section{Results} \label{results_section}

\begin{figure*}
    \centering
    \includegraphics[width=0.99\linewidth]{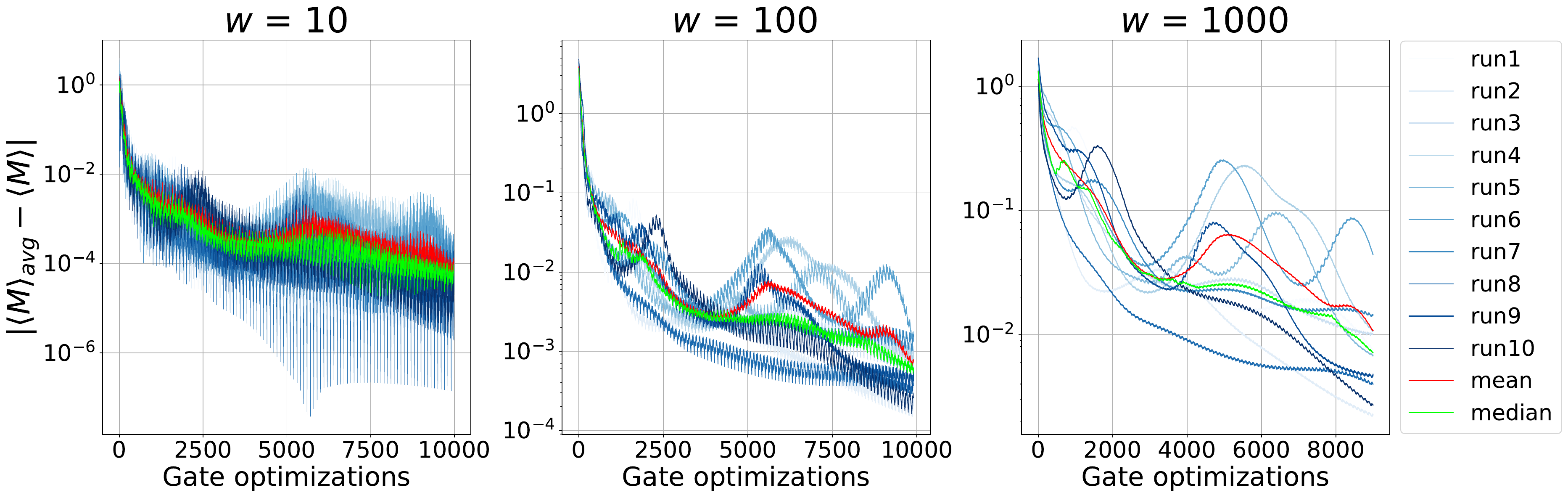}
    \cprotect\caption{Differences between average cost function $\expval{M}_{avg}$ values across gate optimizations compared to the new cost function value $\expval{M}$ on a logarithmic scale. Results are from a 10-qubit Heisenberg model with 10 layers, and \verb|Rotosolve| optimizer was used. Each run is plotted in different shades of blue, the red line is the mean, and the green line represents the median. The averages $\expval{M}_{avg}$ are computed from the latest $w$ gate optimizations. On the left, $w=10$ was used, $w=100$ on the middle plot, and $w=1000$ on the right plot.}
    \label{averages_differences_heisenberg}
\end{figure*}

In this section, we outline the experiments we conducted on various system sizes and complexities. We examine the performance of the hybrid methods proposed for an ideal and noisy device using simulations performed with the PennyLane Python package~\cite{pennylane}. In hybrid methods, we chose the optimizer \verb|Rotosolve| as the less expressive algorithm $\mathcal{A}$ and \verb|FQS| as the more expressive algorithm $\mathcal{B}$ described in Algorithms~\ref{algo1} and~\ref{algo2}, and call the combination of these two optimizers as \verb|RotoFQS| for the convenience. We also provide a performance comparison to gate-wise and iteration hybrid methods from previous work. For gate-wise hybrid, we set the probability $p=0.25$ and 0.5 for the gate to be optimized with \verb|Rotosolve| and $N=2$ for iteration hybrid. That is, we optimize every other iteration with \verb|FQS| and \verb|Rotosolve|. 

\begin{figure}
    \centering
    \includegraphics[width=0.99\linewidth]{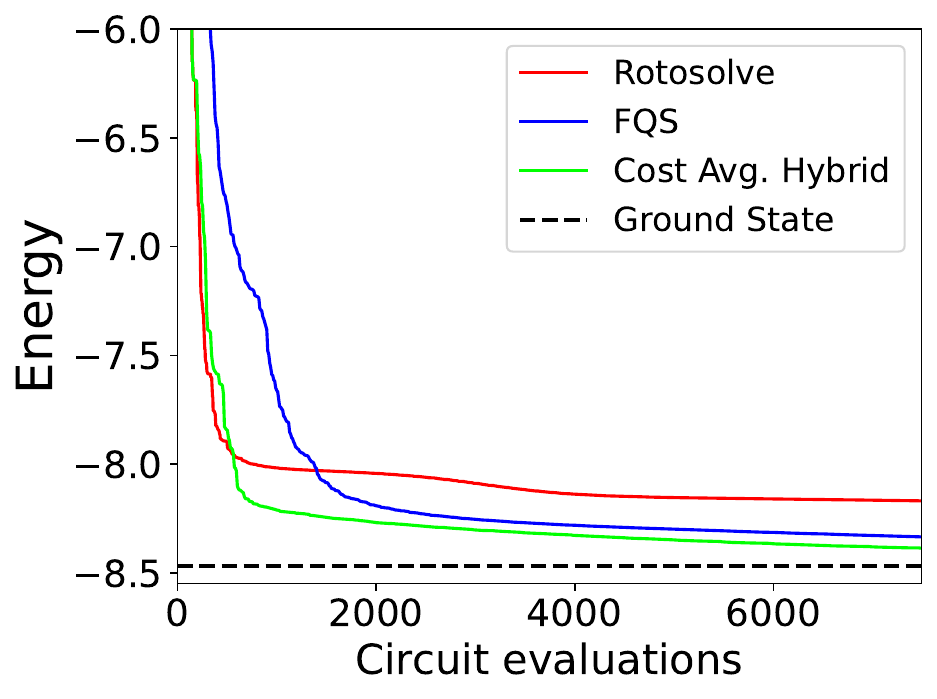}
    \cprotect\caption{Individual runs for \verb|Rotosolve|, \verb|FQS| and cost average hybrid with window length $w=10$ and threshold $E_t=0.05$. All optimizers are initialized with the same parameters, and a 5-qubit Heisenberg model with 10 layers was used.}
    \label{Heisenberg_roto_fqs_same_params}
\end{figure}

First, we present the results for the 10-qubit one-dimensional Heisenberg model with cyclic boundary conditions. Then, we show the results for the 6-qubit Fermi-Hubbard model on a $1\times3$ lattice. Then, we experiment on the scalability of the hybrid methods, where the system size grows in both the number of qubits and layers used in the PQC. For the system, we chose the one-dimensional Heisenberg model and the qubits ranging from 7 to 15, incrementing by two, and set the number of layers to be exactly the number of qubits in the circuit. That is, we set $L = n$ depending on the system size, the number of qubits $n$. Finally, we examine the performance of the optimizer in the fidelity maximization task, where instead of a Hamiltonian, the cost function is a projection operator $\mathcal{P} = - \ket{\phi} \bra{\phi}$ of a random quantum state $\ket{\phi}$, to which we want to optimize the circuit.

In all experiments, we use the ansatz circuit depicted in Fig.~\ref{Ansatz_circuit_image} for all optimizers. For \verb|Rotosolve| we sample each gate $U_d$ from the rotation gates $ \{R_X, R_Y, R_Z\}$ for $d=1,\ldots, Ln$ in the beginning of each run. The optimization of the PQC is done sequentially, starting from the top left gate in the circuit, moving downward afterward to the next gate. Then, all gates in the layer are optimized before moving to the next layer. This process is iterated until all gates in the circuit are optimized once, completing one iteration. Optimization is continued until a given number of iterations or a stopping criterion is reached. In all experiments except in fidelity maximization, each optimizer is run 20 times, and each run consists of 100 iterations for the \verb|Rotosolve| optimizer. To have a fair comparison across different optimizers, we adjust the number of circuit evaluations to be the same amount as for \verb|Rotosolve|. This translates to 50 and 30 iterations for \verb|Fraxis| and \verb|FQS|, respectively, since \verb|Rotosolve| requires 3 circuit evaluations for one gate optimization, \verb|Fraxis| 6 circuit evaluations, and \verb|FQS| 10 circuit evaluations. The hybrid methods are run until the same number of circuit evaluations is reached. 

Before starting the optimization process, we sample the initial parameters uniformly from their corresponding parameter space for each gate. The parameters for \verb|Rotosolve| are sampled from the uniform distribution $(-\pi, \pi]$. The parameters for \verb|Fraxis| are sampled from the spherical uniform distribution, and the parameters for the \verb|FQS| from the spherical uniform distribution in four dimensions. The gate-wise and iteration hybrids are initialized in the same way as described in Ref.~\cite{pankkonen2025improvingvariationalquantumcircuit}.

For the hybrid method that uses cost function averages $\expval{M}_{avg}$, we illustrate the behavior of the computed values of $|\expval{M}_{avg} - \expval{M}|$ in Fig.~\ref{averages_differences_heisenberg} as a function of gate optimizations. Here we have plotted 10 individual runs of \verb|Rotosolve| on the one-dimensional Heisenberg model with 10 qubits and layers, and an ideal quantum device was used in simulation. On the left sub-figure, we used the window size $w=10$, $w=100$ in the middle, and $w=1000$ on the right. In each sub-figure, the individual runs are shown in different shades of blue, and the mean and median across all runs are shown in red and green, respectively. We remark that when the window size $w$ varies, the log-scale of the vertical axis changes accordingly. With a smaller $w$, the absolute difference of averages from the new value of the cost function oscillates rapidly and with high amplitude compared to the scale. When the window of the cost function average increases, the oscillation dampens a bit, and a clearer trend of each run emerges. When $w=1000$, we can see that the oscillation becomes nearly zero, and a clear trend can be seen. Also, the results in Fig.~\ref{averages_differences_heisenberg} imply that when choosing parameters for $E_t$ and $w$, we can keep either one fixed and only try different values for the other. In the following experiments, we keep $E_t$ fixed and do the experiments by trying different values for $w$. In Fig.~\ref{Heisenberg_roto_fqs_same_params}, we illustrate how individual runs of \verb|Rotosolve|, \verb|FQS|, and the cost average hybrid behave for a 5-qubit Heisenberg Hamiltonian with 10 layers using an ideal quantum device. Here we set $w=10$ and $E_t=0.05$ for the cost average hybrid. In Fig.~\ref{Heisenberg_roto_fqs_same_params}, all optimizers are initialized with the same parameters. First, we sample gates $R_i \in \{R_X, R_Y, R_Z\}$ randomly and the parameters $\theta_i \in (-\pi,\pi]$ uniformly. For \verb|FQS|, the gates $R_i$ with given parameters $\theta_i$ are transformed into a quaternion representation in Ref.~\cite{fqs}. As we simulate the runs with an ideal quantum device, the cost average hybrid follows the \verb|Rotosolve| until the switching criterion is met. After that, the cost-average hybrid has an advantage over regular \verb|FQS| as it has more circuit evaluations left to use and thus achieves better convergence than the regular \verb|FQS|. 

In Appendix~\ref{appendix_window_lengths}, we provide additional results for \verb|Rotosolve| with varying numbers of shots used with the noisy device of the Heisenberg model. We also provide simulation results and illustrations for the \verb|Fraxis| and \verb|FQS| optimizers with an ideal quantum device.

\subsection{Heisenberg model}

In this section, we consider the one-dimensional Heisenberg model~\cite{heisenberg_model} with cyclic boundary conditions and an external magnetic field along the Z-axis. The corresponding Hamiltonian is defined as 
\begin{equation}\label{heisenberg_ham_eq}
    H  = \mathcal{J} \sum_{i=1}^n(X_i X_{i+1} + Y_i Y_{i+1} + Z_i Z_{i+1})+ h \sum_{i=1}^n Z_i,
\end{equation}
where $\mathcal{J}$ is the strength of the spin interaction and $h$ describes the strength of an external magnetic field along the Z-axis. In this work, we set $\mathcal{J} = h = 1$ in the following experiments.

\begin{figure}
    \centering
    \includegraphics[width=0.99\linewidth]{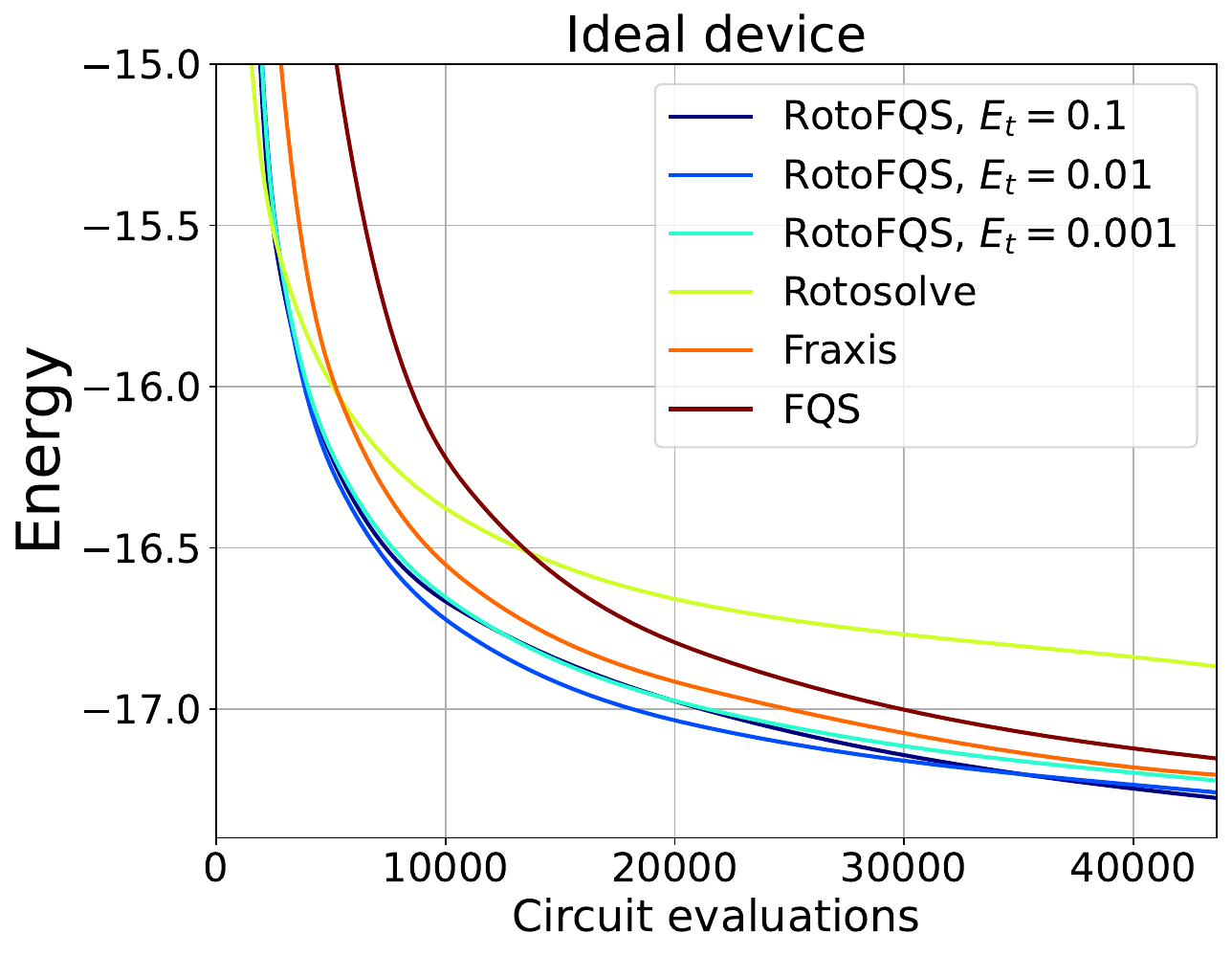}
    \includegraphics[width=0.99\linewidth]{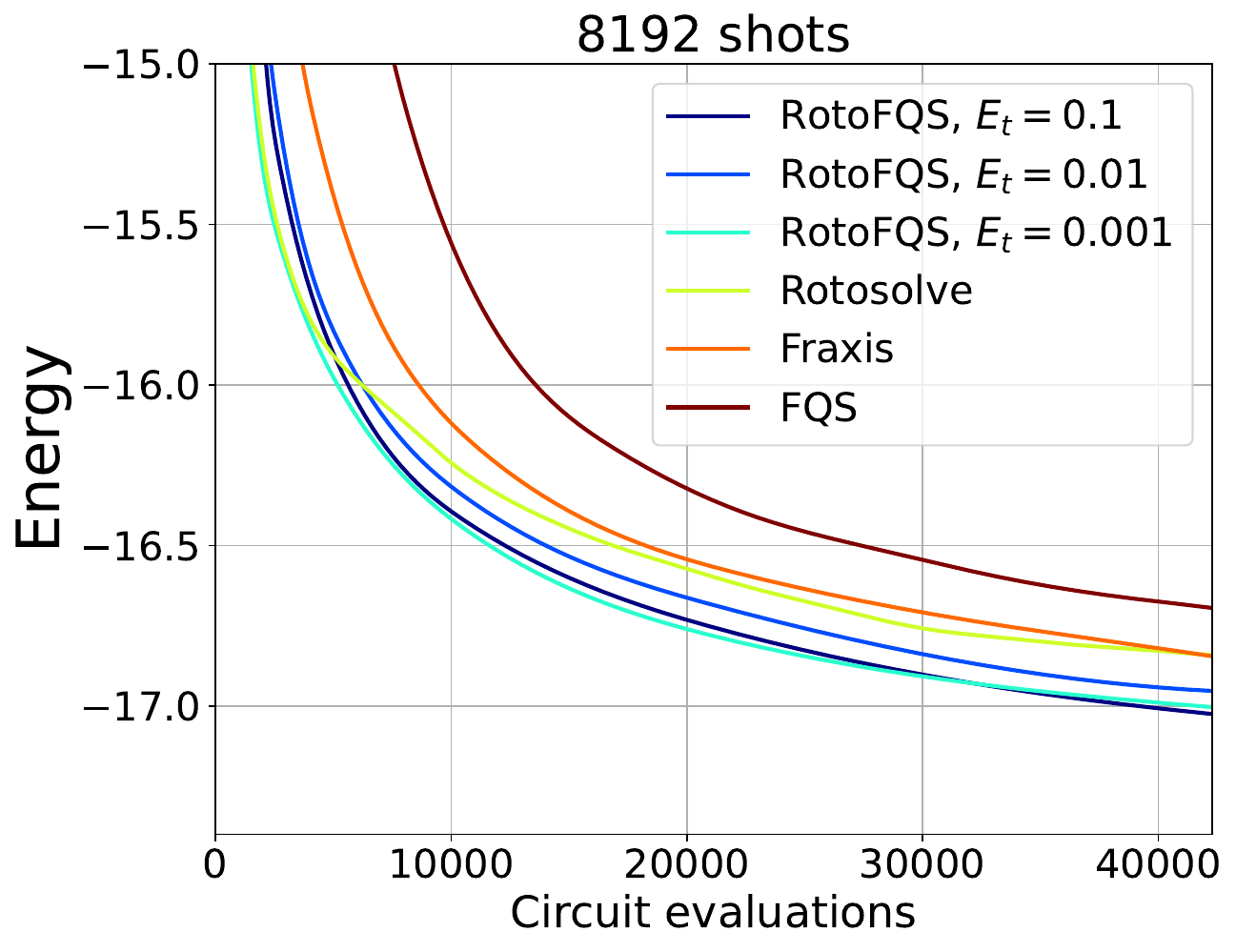}
    \cprotect\caption{Results for one-dimensional 10-qubit Heisenberg model with different optimizers \verb|Rotosolve| (yellow), \verb|Fraxis| (orange), \verb|FQS| (dark red), and \verb|RotoFQS| with thresholds $E_t = $ 0.1 (dark blue), 0.01 (blue), and 0.1 (cyan). The patience of the hybrid algorithm was set to $P=10$. Each line represents the mean across the 20 runs.}
    \label{hybrid_heisenberg_results}
\end{figure}

We provide results for the 10-qubit Heisenberg model with 15 layers. We used thresholds $E_t = 0.1, 0.01$ and $0.001$ for \verb|RotoFQS| with patience set to $P=10$. A total of 20 runs were performed for each optimizer. To ensure a fair comparison of the optimizers' performance, the number of iterations was set to 100 for \verb|Rotosolve|, 50 for \verb|Fraxis|, and 30 for \verb|FQS|. The \verb|RotoFQS| and other hybrids were run until the set number of circuit evaluations was reached. In our experiment, we simulated an ideal device and a noisy device with 8192 shots. We also simulated the performance of the cost function average hybrid method with an ideal quantum device and window lengths set to $w=10$, $100$, and $1000$. The threshold for switching the optimizer was set to $ E_t = 0.01$. Finally, we compare the performance of cost function-based hybrid methods to gate-wise and iteration hybrids.

\begin{figure*}
    \centering
    \includegraphics[width=0.99\linewidth]{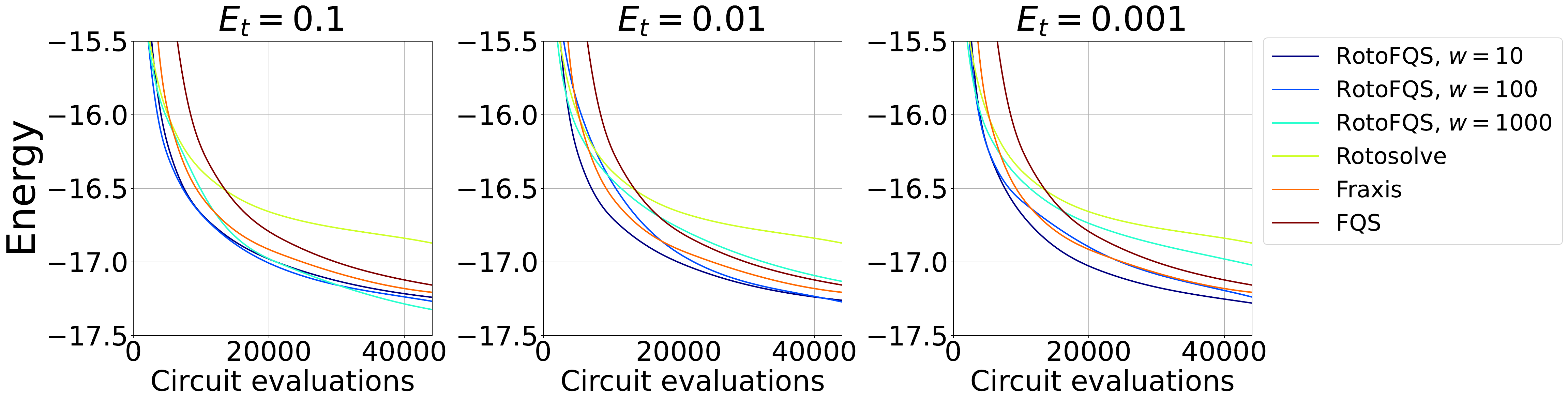}
    \cprotect\caption{Results for one-dimensional 10-qubit Heisenberg model with 15 layers. Each line represents a different optimizer: \verb|Rotosolve| (yellow), \verb|Fraxis| (orange), \verb|FQS| (dark red), and \verb|RotoFQS| with window lengths $w = $ 10 (dark blue), 100 (blue), and 1000 (cyan). The switching threshold for the optimizers was set to $E_t=0.1$ (left), 0.01 (middle), and 0.001 (right). Each line represents the mean across the 20 runs.}
    \label{hybrid_heisenberg_results_cost_average_func}
\end{figure*}

\begin{figure}
    \centering
    \includegraphics[width=0.99\linewidth]{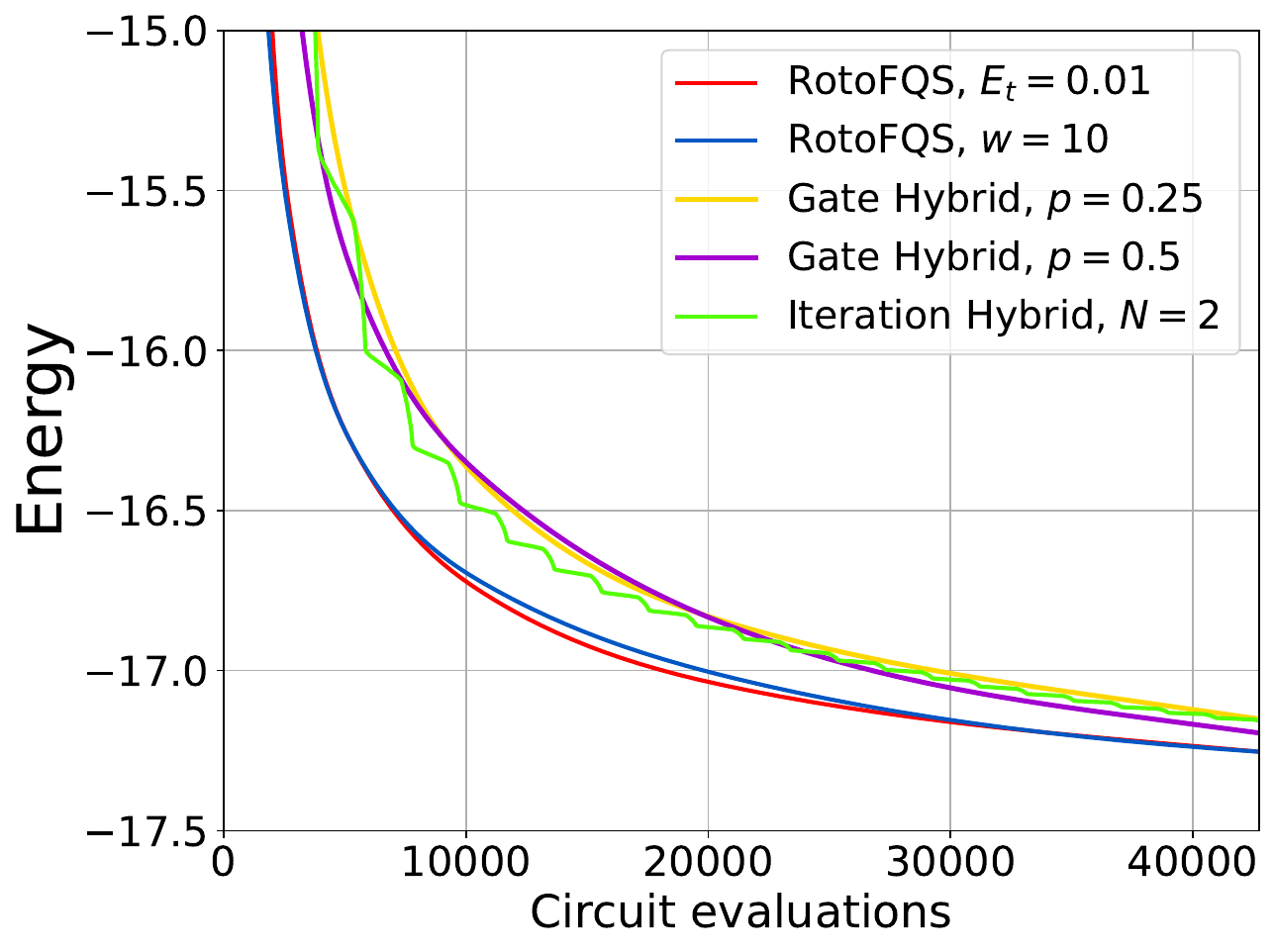}
    \cprotect\caption{A comparison of all hybrid methods: early stopping (red), cost-average (blue), gate-wise with $p=0.25$ (yellow), $p=0.5$ (purple), and iteration hybrid (green) with $N=2$ for a one-dimensional 10-qubit Heisenberg model with 15 layers. The patience of the early stopping hybrid algorithm was set to $P=10$, and the window length $w=10$ was set for the cost-average hybrid (blue), and in both $E_t=0.01$. Each line represents the mean across the 20 runs.}
    \label{ALL_HYBRIDS_heisenberg_results}
\end{figure}

The results for the early stopping hybrid are fully shown in Fig.~\ref{hybrid_heisenberg_results}. With the ideal device, regardless of the threshold values, \verb|RotoFQS| achieves faster and better convergence than any other optimizer. This is emphasized when using a noisy device with 8192 shots. With the ideal device \verb|Fraxis| comes close to hybrid methods, \verb|FQS| following tightly behind, and the \verb|Rotosolve| has the worst performance. When we use a noisy device, the performance of the \verb|RotoFQS| becomes more distinct when compared to other optimizers, namely, achieving relatively better convergence than with the ideal device. In addition, \verb|Rotosolve| and \verb|Fraxis| have equal convergence, but \verb|Rotosolve| does it faster. \verb|FQS| on the other hand, performs the worst when a noisy device is used. This can be explained that \verb|FQS| requires 10 circuit evaluations in order to optimize a single gate, while \verb|Rotsolve| and \verb|Fraxis| require 3 and 6, respectively. 

The results for the cost-average-based hybrid algorithm are shown in Fig.~\ref{hybrid_heisenberg_results_cost_average_func}. Each subplot represents a different window of the cost function average, which we use to determine switching for the \verb|RotoFQS|. With $E_t = 0.1$, all the average methods of the cost function with different window lengths achieve better convergence than other optimizers, where the best is the window length $w=1000$. However, when we narrow the threshold $E_t$ and the longer the cost function average window is, the worse it tends to perform. On the other hand, narrowing the threshold improves the performance for smaller window lengths, as seen for $w=10$ when comparing the left and right subplots of Fig.~\ref{hybrid_heisenberg_results_cost_average_func}.

Finally, we compare the performance of cost-based hybrids with the gate-wise and iteration hybrids of the previous work in Fig.~\ref{ALL_HYBRIDS_heisenberg_results}. For cost-based hybrids, we set $w=10$ for the cost-average hybrid and $P=10$ for early stopping, and for both, the threshold $E_t$ is set to 0.01. The cost-based hybrids of \verb|RotoFQS| have a better and faster convergence than gate or iteration hybrids. The zigzag pattern of the iteration hybrid method comes from every other iteration being \verb|FQS| and every other \verb|Rotosolve|. The longer and horizontal parts of the mean come from \verb|FQS| iterations. The shorter and vertical parts are optimization iterations done with \verb|Rotosolve|. This drastic decrease in the cost function is explained by the fact that the gate axes and angles are optimized simultaneously by \verb|FQS|, and subsequently, \verb|Rotosolve| easily finds better minima with a lower cost. That is, with already optimized axis of rotation and angle, the \verb|Rotosolve| can find much better minima after the \verb|FQS| optimization. Overall, the gate-wise hybrids with $p=0.25$ and $p=0.5$ perform equally well alongside the iteration hybrid. 

\begin{figure*}
    \centering
    \includegraphics[width=0.99\linewidth]{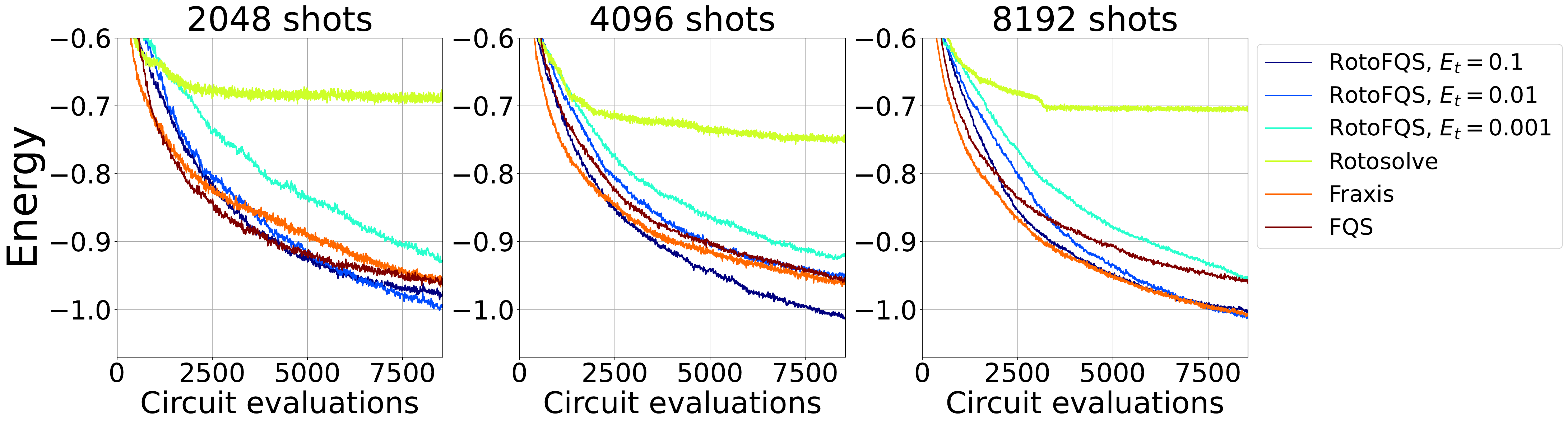}
    \includegraphics[width=0.99\linewidth]{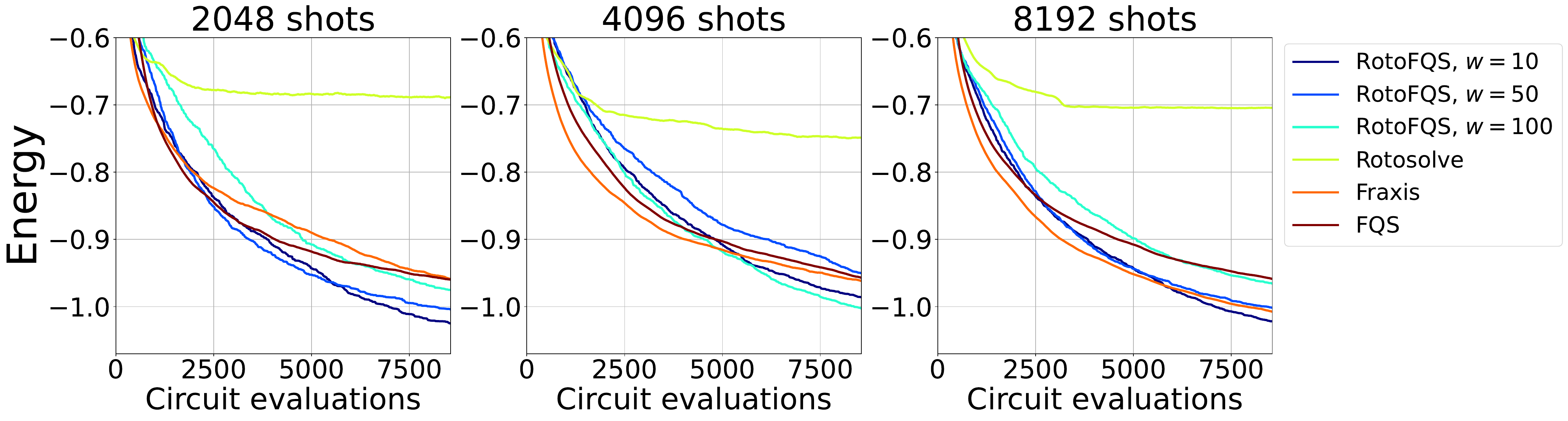}
    \cprotect\caption{Results for one-dimensional 6-qubit Fermi-Hubbard model with 5 layers on a $1\times 3$ lattice. Optimizers \verb|Rotosolve| (yellow), \verb|Fraxis| (orange), \verb|FQS| (dark red). In the top row, early stopping for \verb|RotoFQS| with thresholds $E_t = $ 0.1 (dark blue), 0.01 (blue), and 0.001 (cyan) was used, and the patience of the hybrid algorithm was set to $P=10$. In the bottom row, $E_t$ was set to 0.01 with $w=10$ (dark blue), $w=50$ (blue), and $w=100$ (cyan). In the left column, 2048 shots were used to approximate each Hamiltonian term, 4096 in the middle column, and 8192 in the right column. Each line represents the mean across the 20 runs.}
    \label{hybrid_early_stopping_hubbard_results_noisy}
\end{figure*}

\begin{figure}
    \centering
    \includegraphics[width=0.99\linewidth]{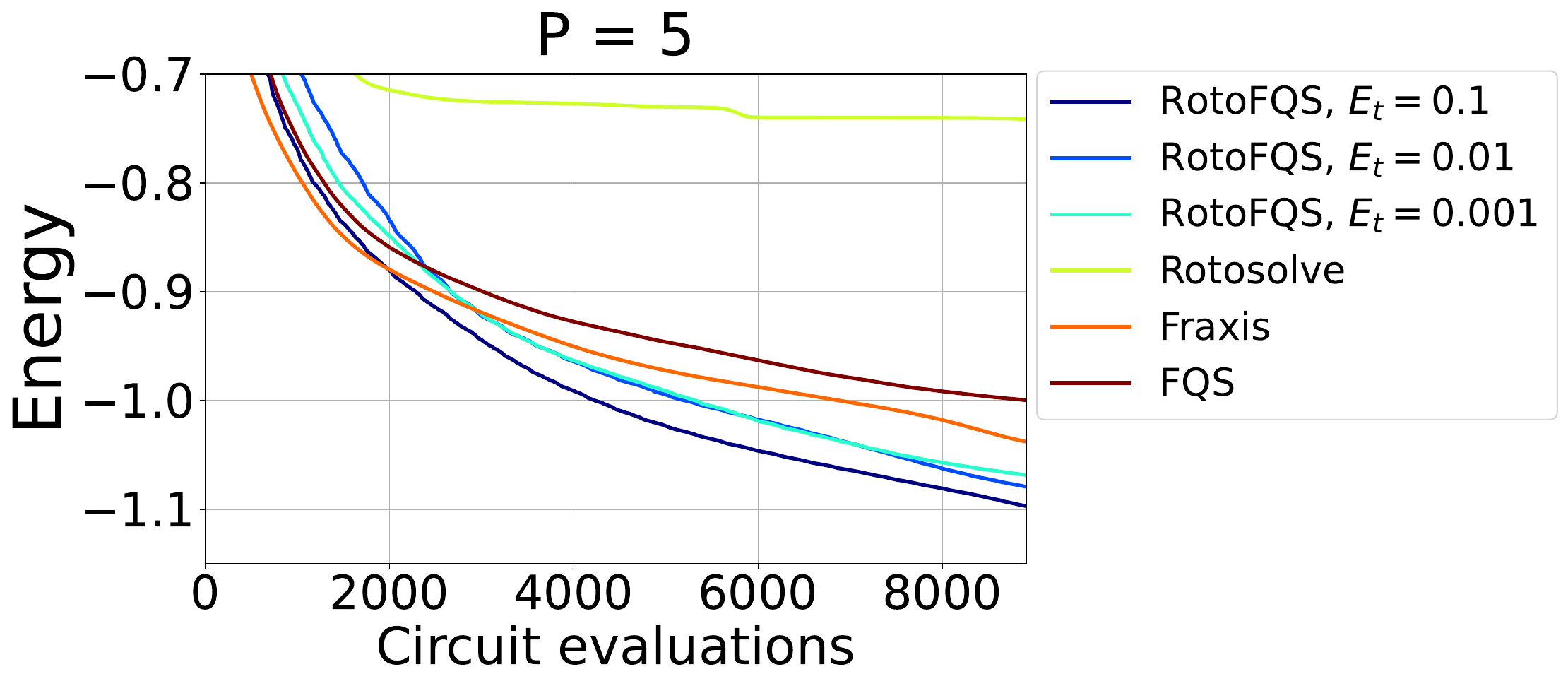}
    \includegraphics[width=0.99\linewidth]{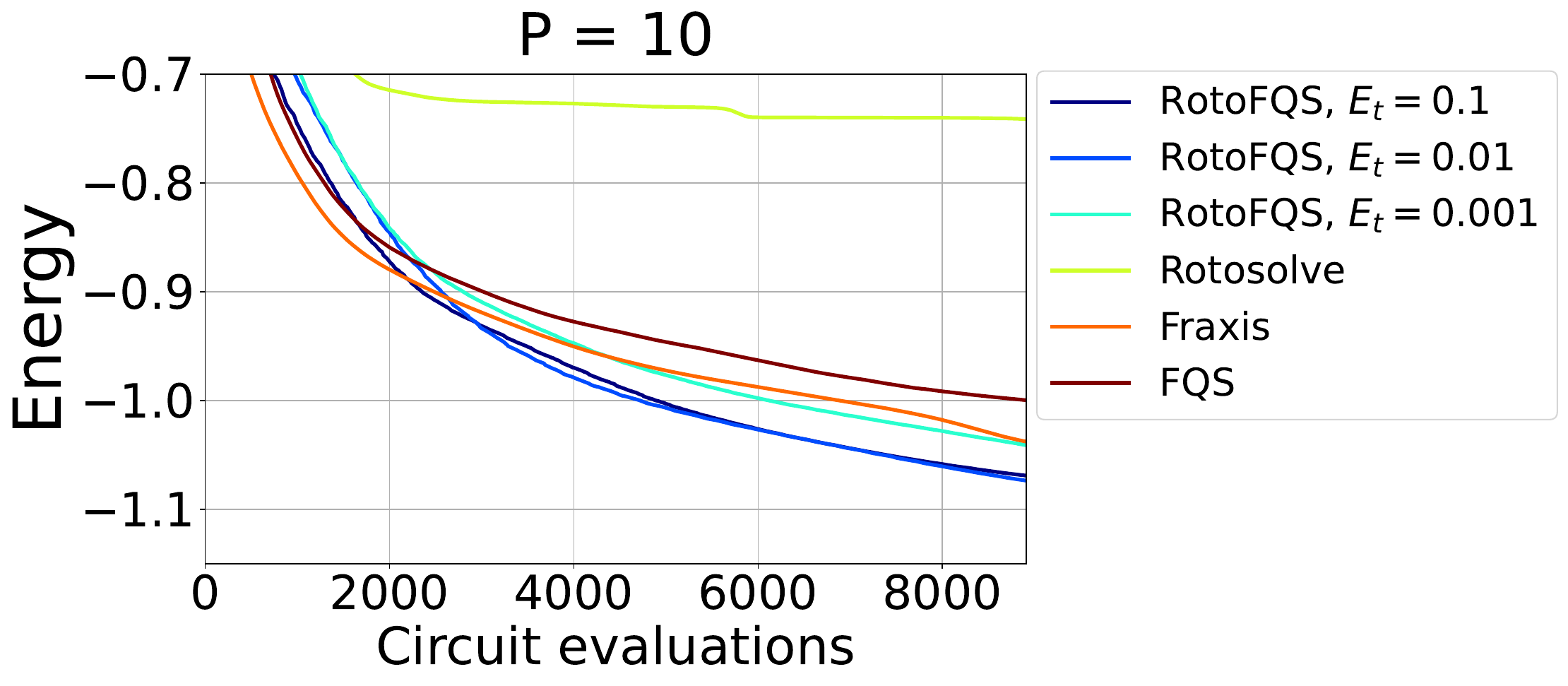}
    \cprotect\caption{Results for 6-qubit Fermi-Hubbard model with 5 layers on a $1\times 3$ lattice. Optimizers \verb|Rotosolve| (yellow), \verb|Fraxis| (orange), \verb|FQS| (dark red), and \verb|RotoFQS| with thresholds $E_t = $ 0.1 (dark blue), 0.01 (blue), and 0.001 (cyan) were used. The patience of the hybrid algorithm was set to $P=5$ (top) and $P=10$ (bottom). Each line represents the mean across the 20 runs.}
    \label{hybrid_early_stopping_hubbard_results_ideal}
\end{figure} 

\begin{figure*}
    \centering
    \includegraphics[width=0.99\linewidth]{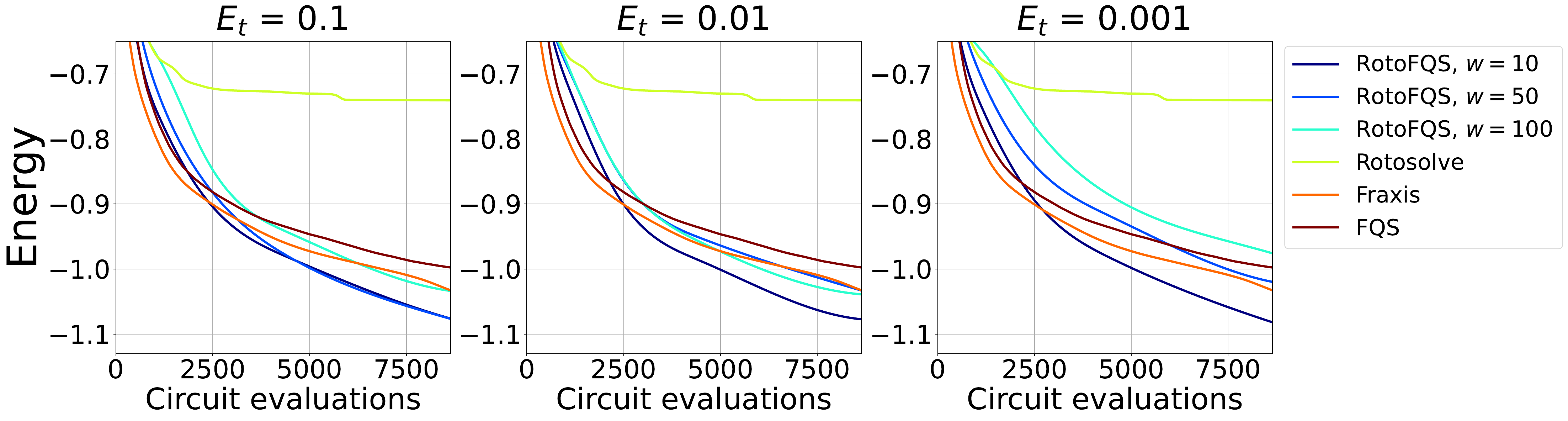}
    \cprotect\caption{Results for 6-qubit Fermi-Hubbard model with 5 layers on a $1\times 3$ lattice. Optimizers \verb|Rotosolve| (yellow), \verb|Fraxis| (orange), \verb|FQS| (dark red), and \verb|RotoFQS| with window lengths $w = 10$ (dark blue), 50 (blue), and 100 (cyan) were used. The switching threshold of the hybrid algorithm was set to $E_t = 0.1$ (left), 0.01 (middle), and 0.001 (right). Each line represents the mean across the 20 runs.}
    \label{hybrid_cost_average_hubbard_results}
\end{figure*}

\begin{figure}
    \centering
    \includegraphics[width=0.99\linewidth]{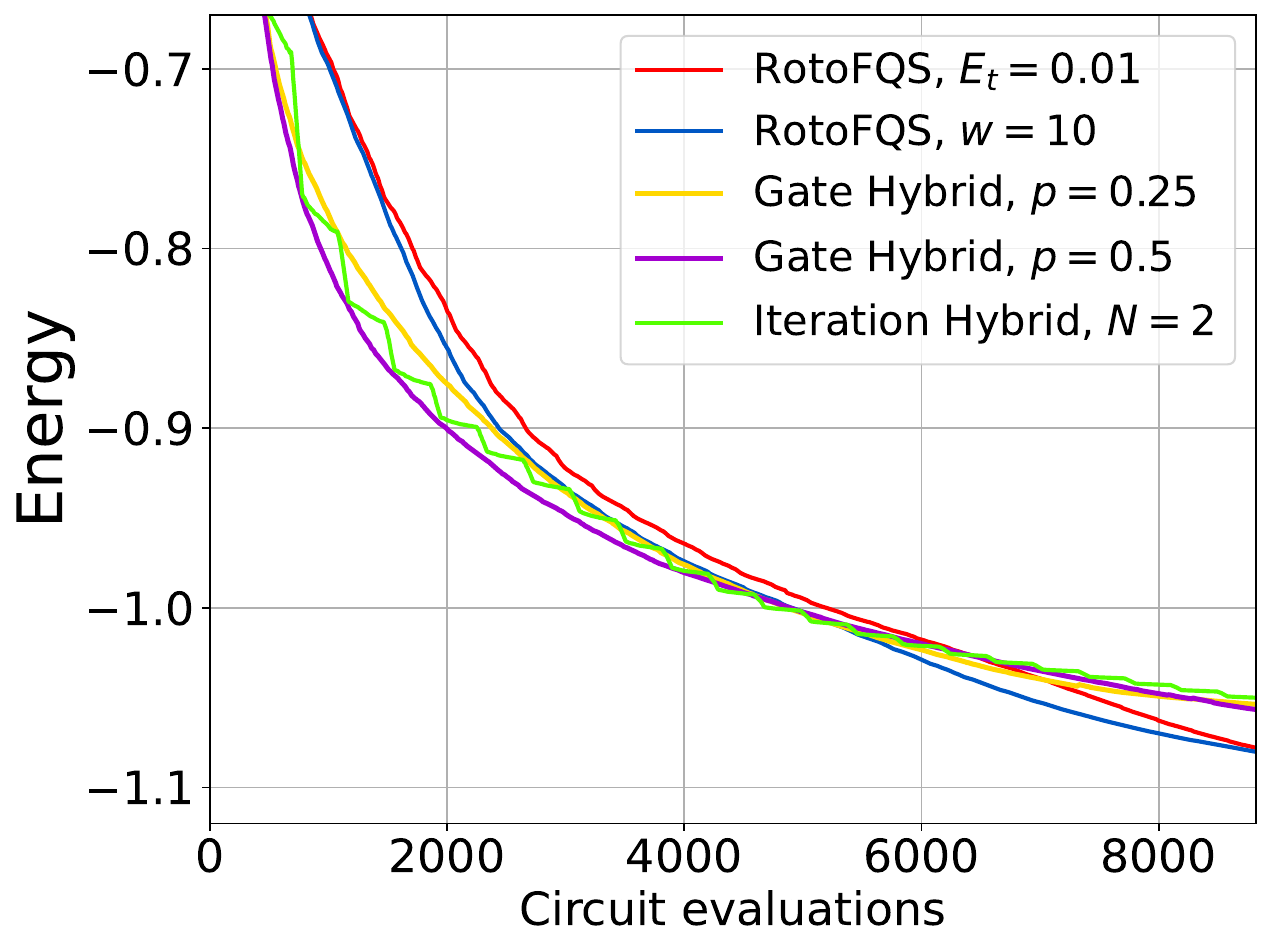}
    \cprotect\caption{A comparison of all hybrid methods: early stopping (red), cost-average (blue), gate-wise with $p=0.25$ (yellow), $p=0.5$ (purple), and iteration hybrid (green) with $N=2$ for a one-dimensional 6-qubit Fermi-Hubbard model with 5 layers on a $1\times 3$ lattice. The patience of the early stopping hybrid algorithm was set to $P=5$, and the window length $w=10$ was set for the cost-average hybrid (blue), and in both $E_t=0.01$. Each line represents the mean across the 20 runs.}
    \label{ALL_HYBRIDS_fermihubbard_results}
\end{figure}

\subsection{Fermi-Hubbard Model}

In this section, we present our results for the Fermi-Hubbard model \cite{fermi_hubbard_model}, which describes how fermions interact in a lattice. The Fermi-Hubbard model Hamiltonian is defined as follows \cite{hubbard_hamiltonian}

\begin{equation}
    H = -t \sum_{<i,j>,\sigma} (\hat{c}_{i,\sigma}^\dagger \hat{c}_{j,\sigma} + \text{h.c.}) + U\sum_i \hat{n}_{i,\uparrow}\hat{n}_{i,\downarrow},
\end{equation}
where the first term is the kinetic term, $<i,j>$ denotes the neighboring lattice sites and $t$ is the tunneling matrix. $\hat{c}_{i,\sigma}^\dagger$ and  $\hat{c}_{j,\sigma}$ denote the fermionic creation and annihilation operators, respectively, and h.c. stands for the hermitian conjugate. Furthermore, the operators $\hat{c}_{j,\sigma}^\dagger$ and $\hat{c}_{j,\sigma}$ correspond to adding or removing a fermion with the spin state $\sigma$ on the site $j$. The last term containing the number operators $\hat{n}_{i,\uparrow}$ and $\hat{n}_{i,\downarrow}$ is the potential term. These number operators describe the number of particles in the up- or down states in the site $i$. The $U$ denotes the strength of the interaction.

We aim to determine the ground state of this model for a $1\times3$ lattice with a 6-qubit system, setting $t = U = 0.5$ in this work. The corresponding Hamiltonian for the parameters $t$ and $U$ is extracted by using the Pennylane Python package~\cite{pennylane}, and with applying Jordan-Wigner mapping~\cite{jordan_wigner} to the fermionic creation and annihilation operators $\hat{c}_{i,\sigma}^\dagger$ and $\hat{c}_{j,\sigma}$, respectively. The ground state of the Hamiltonian is approximately $E_g = -1.25$, which is the lowest eigenvalue computed by using the Pennylanes \verb|qml.eigvals| function.

Next, we show the results for the early stopping and cost function average hybrid methods for \verb|RotoFQS|. A total of 20 runs were performed in all subsequent experiments, and with the same number of iterations as in the previous subsection for the Heisenberg model.

In Fig.~\ref{hybrid_early_stopping_hubbard_results_noisy}, we provide results for a 6-qubit Fermi-Hubbard Hamiltonian with 5 layers on a noisy device, where we simulate noise with 2048, 4096, and 8192 shots. We compare \verb|Rotosolve|, \verb|Fraxis|, and \verb|FQS| to early stopping hybrid and cost-average-based hybrid methods. With the fewest shots, which demonstrates the noisiest setup for the optimization, the early stopping hybrid of \verb|RotoFQS| with threshold $E_t = 0.01$ is the best, followed by $E_t=0.1$ and then \verb|Fraxis|. Interestingly, \verb|Fraxis| and \verb|FQS| have faster initial convergence, but \verb|RotoFQS| manages to keep the convergence going and not plateau. This can also be observed with 4096 shots, but now only \verb|Fraxis| has the better initial convergence. Again, in the end, the \verb|RotoFQS| is better than the rest where $E_t=0.01$. With 8192 shots, the \verb|RotoFQS| with early stopping achieves equal convergence to \verb|Fraxis| with $E_t = 0.1$ and 0.01. Due to a noisy device, \verb|RotoFQS| with early stopping and small threshold $E_t=0.001$ performs the worst after \verb|Rotosolve|. The noise from the measurements affects the optimization, and the measured cost function does not necessarily fall in the range of the threshold $E_t$ compared to the previous value, even though on the ideal device, the differences can be minuscule and easily be below the threshold value $E_t$. For the cost-average-based hybrid, we used a switching threshold $E_t = 0.01$ for all window lengths $w=10, 50$, and 100. For all levels of noise, $w=10$ is comparably the best when considering the convergence of each optimizer. With 2048 and 8192 shots, $w=10$ is the best, followed by $w=50$ and then \verb|Fraxis|. However, with the intermediate noise level of 4096 shots, $w=100$ achieves the best convergence, and then $w=10$. When the shots are set to 2048 or 8192, the higher the window length is, the worse the hybrid method tends to perform. Additionally, when comparing the results of both hybrid methods with each other, they seem to perform more or less equally well across different numbers of shots to simulate the noise in the quantum devices. That is, hybrid methods benefit from the noise, but other optimizers, \verb|Fraxis|, especially begin to converge toward the performance of \verb|RotoFQS|, regardless of the hybrid method with which it is used.

Next, with the ideal quantum device, we simulated an early stopping hybrid with patience set to $P=5$ and $P=10$ and threshold values $E_t = 0.1$, 0.01, and 0.001. The results are fully shown in Fig.~\ref{hybrid_early_stopping_hubbard_results_ideal}. When the patience is set to $P=5$, the early stopping hybrid version of \verb|RotoFQS| with $E_t = 0.1$ is the best, followed by $E_t = 0.01$ and $E_t = 0.001$, respectively. In both sub-figures, the standalone optimizers \verb|Rotosolve|, \verb|FQS|, and \verb|Fraxis| are outperformed by the \verb|RotoFQS| regardless of the threshold value $E_t$. By closer examining the \verb|RotoFQS|, we see that with a lower patience value, we obtain better results.

Then, we simulated the cost-average-based hybrid method with an ideal quantum device and set the switching thresholds $E_t = 0.1, 0.01$, and 0.001. For each $E_t$, we tested window lengths $w=10, 50$, and 100. The results are fully shown in Fig.~\ref{hybrid_cost_average_hubbard_results}. With the ideal quantum device, the results are much more favorable for the cost-average-based hybrid than in the noisy devices. Regardless of the switching threshold $E_t$, the window length $w=10$ performs the best in terms of the convergence across 20 runs. After that, either one of the hybrids is next, depending on the $E_t$ or \verb|Fraxis|. When we decrease the switching threshold $E_t$ to smaller and smaller values, the worse hybrids with $w=50$ and $w=100$ will perform. The higher complexity of the Hamiltonian can explain this compared to the Heisenberg model, as well as the fact that cost function values need to plateau faster so that the algorithm is switched to more expressive \verb|FQS|. 

Finally, we examine the performance of all hybrid methods in Fig.~\ref{ALL_HYBRIDS_fermihubbard_results}. For \verb|RotoFQS| we set $w=10$ for the cost-average hybrid and patience $P=5$ for the early stopping hybrid. In both, the threshold $E_t$ is set to 0.01. Iteration and gate-wise hybrids all exhibit similar convergence when compared to each other. At the beginning of the optimization, they have a bit faster initial convergence, but the \verb|RotoFQS| hybrids exceed them in the end.

\begin{figure*}
    \centering
    \includegraphics[width=0.99\linewidth]{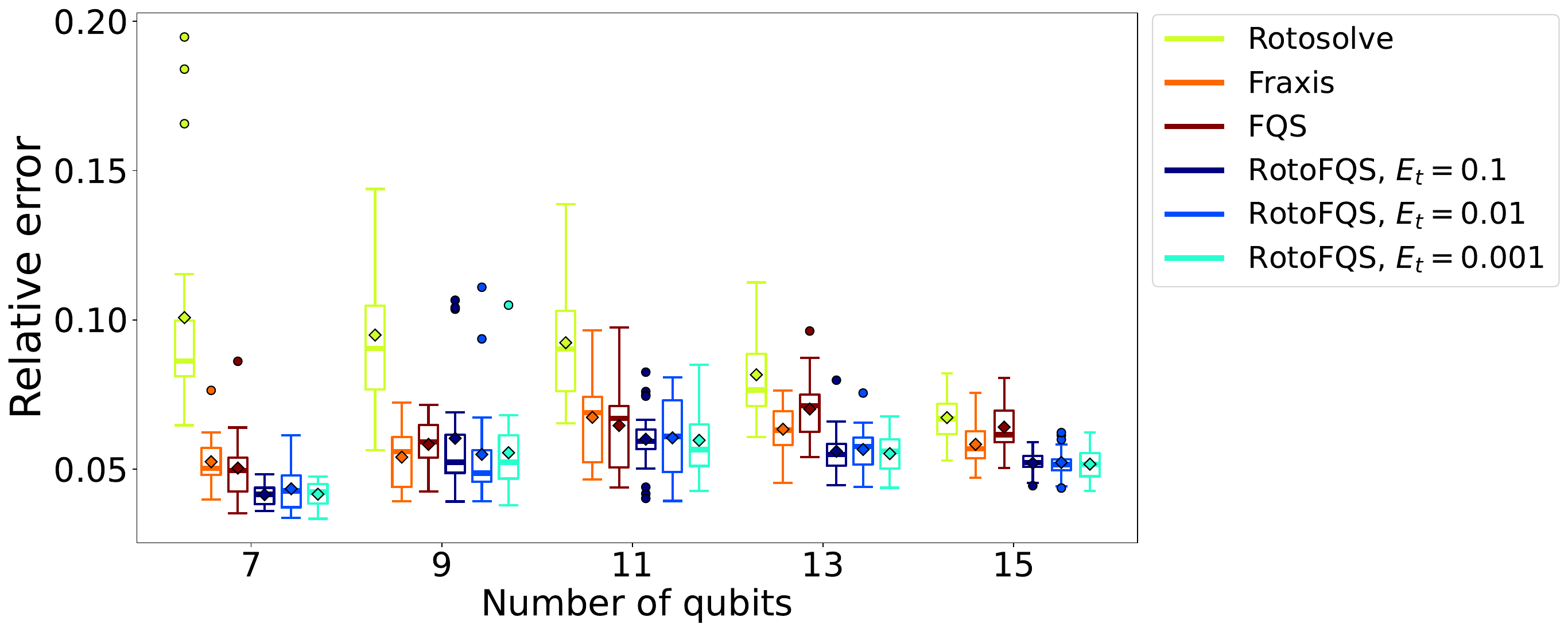}
    \includegraphics[width=0.99\linewidth]{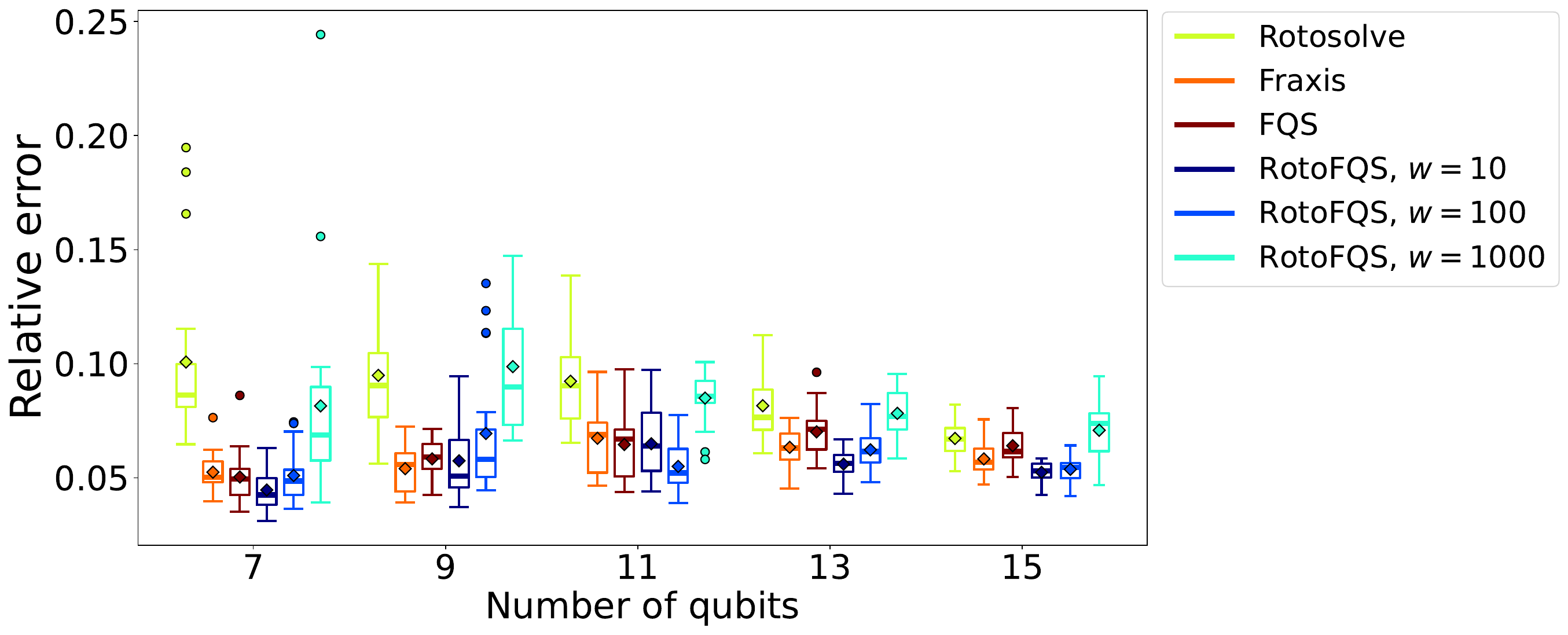}
    \includegraphics[width=0.99\linewidth]{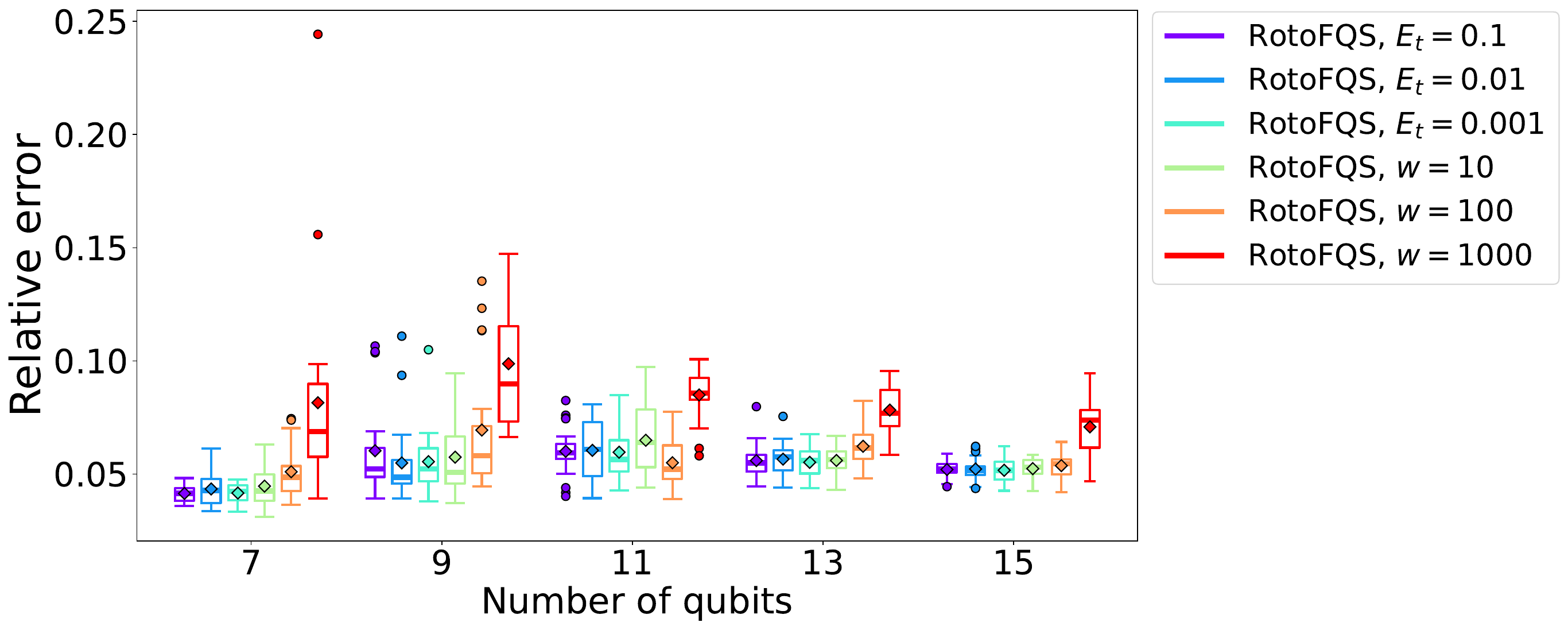}
    \cprotect\caption{Relative errors for the Heisenberg model for qubits ranging from 7 to 15, incrementing by 2. The \verb|Rotosolve| (yellow), \verb|Fraxis| (orange), and \verb|FQS| (dark red) optimizers are compared to the early stopping hybrid method (top) and cost-average-based hybrid method (middle). Both hybrid methods are compared to each other in the bottom sub-figure. The vertical axis denotes the relative error from the ground state, and the horizontal axis is the number of qubits. The number of layers was set to $L=n$ for each system size of $n$ qubits. In all legends for \verb|RotoFQS|, $E_t$ denotes the threshold of the early stopping hybrid with patience $P=5$, and $w$ denotes the window size used for the cost-average hybrid while setting the threshold to $E_t=0.01$.}
    \label{hybrid_scalability_results_early_stop}
\end{figure*}

\begin{figure*}
    \centering
    \includegraphics[width=0.99\linewidth]{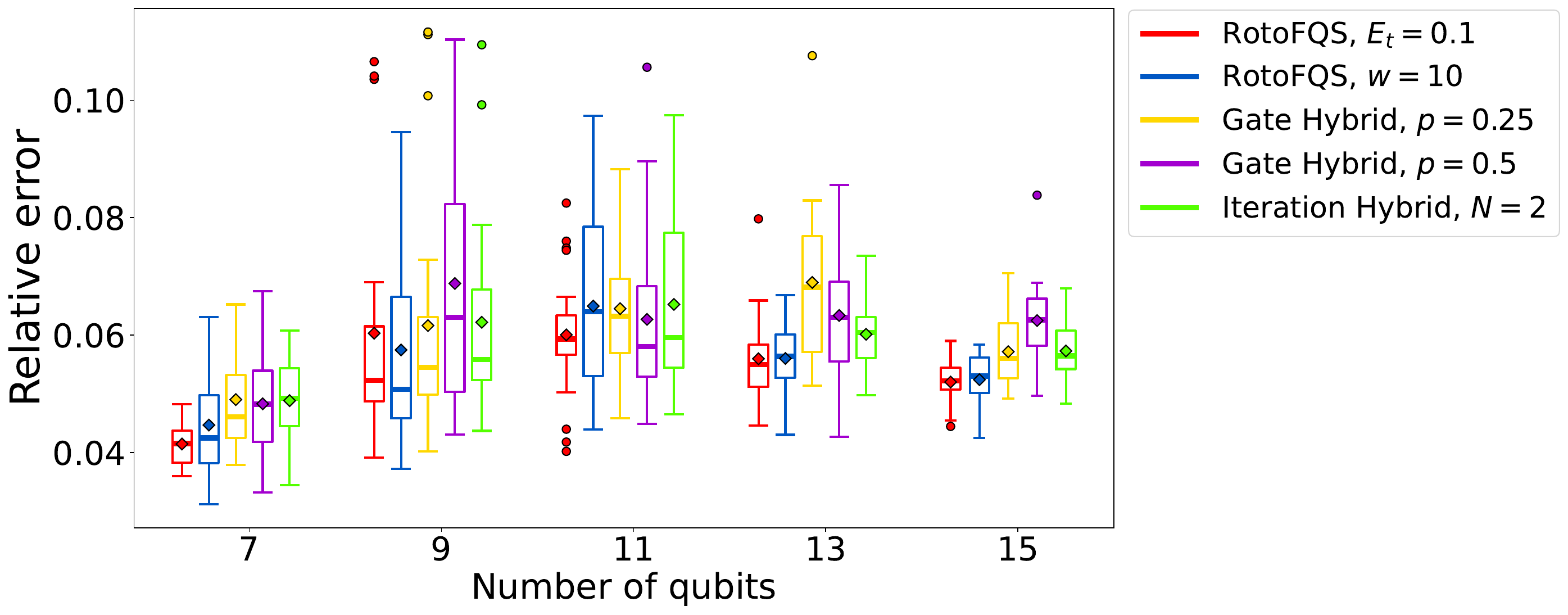}
    \cprotect\caption{Relative errors for all hybrid methods: early stopping (red), cost-average (blue), gate-wise with $p=0.25$ (yellow), $p=0.5$ (purple), and iteration hybrid (green) with $N=2$ for a one-dimensional Heisenberg model, with the number of layers was set to $L=n$ for each system size of $n$ qubits. The patience of the early stopping hybrid algorithm was set to $P=5$ with threshold $E_t=0.1$, and the window length $w=10$ was set for the cost-average hybrid (blue) with threshold $E_t=0.01$.}
    \label{ALL_HYBRIDS_scalability_results}
\end{figure*}
\subsection{Scalability}

We investigate the scalability of our proposed hybrid methods, early stopping, and cost function average for \verb|RotoFQS| and compare it to the base version of \verb|Rotosolve|, \verb|Fraxis|, and \verb|FQS| across the system sizes. We chose the Heisenberg model as it has been highly used in the literature regarding quantum computation~\cite{Ostaszewski_2021, fraxis, zhang2022escaping, jattana2022assessment, shang2023schrodinger, stein2022eqc, wiersema2020exploring, verdon2019quantum, kokail2021quantum, bosse2022probing, liu2019variational}. We scale the Heisenberg model Hamiltonian of Eq.~(\ref{heisenberg_ham_eq}) from 7 to 15 qubits, incrementing by 2. For each system size, we use $L=n$ layers for the ansatz circuit, depicted in Fig.~\ref{Ansatz_circuit_image}. As in the previous sections, we execute 20 runs for each optimizer, and for each run, we use 100 iterations for \verb|Rotosolve|, 50 for \verb|Fraxis|, and 30 for \verb|FQS|, respectively. Both hybrid methods for \verb|RotoFQS| are initialized in the same way as \verb|Rotosolve| and are run until the set number of circuit evaluations is reached. The patience for the early stopping hybrid was set to $P=5$ in all system sizes with thresholds $E_t = 0.1, 0.01, 0.001$. For the cost-average-based hybrid, we used window lengths $w=10, 100, 1000$ with a switching threshold of $E_t=0.01$. We compare the relative error to the ground state for each system size. The ground states of the systems are computed by using the QuSpin Python package~\cite{Quspin_package}. 

In Fig.~\ref{hybrid_scalability_results_early_stop}, we provide a full comparison of both hybrid algorithms, early stopping (top sub-figure) and cost-average-based hybrid (middle sub-figure) to \verb|Rotosolve|, \verb|Fraxis|, and \verb|FQS|. In addition, we also make a full comparison of both hybrid algorithms to each other in the bottom sub-figure of Fig.~\ref{hybrid_scalability_results_early_stop}. The boxes in all sub-figures span from the first quartile (Q1) to the third quartile (Q3); the horizontal bar within each box indicates the median, the diamond marker denotes the mean, and the whiskers represent values within 1.5 times the interquartile range (IQR).

The early stopping hybrid, regardless of the threshold $E_t$, achieves the best mean and median values across the system sizes used compared to \verb|Rotosolve|, \verb|Fraxis|, and \verb|FQS|. This becomes increasingly notable as the number of qubits and layers in the ansatz circuit increases, demonstrating good scalability. In addition to better mean and median values, the early stopping hybrids have a much better concentration of runs with a higher number of qubits, making it more reliable to obtain good convergence for the given ansatz circuit and computational resources (circuit evaluations). When the cost-average-based hybrid is used, the results differ a bit compared to early stopping. With $w=10$, the best results are achieved in terms of median, mean, and overall narrow concentration of the runs. However, when the window length is increased, the performance increasingly becomes closer to \verb|Rotosolve| because the switch between \verb|Rotosolve| and \verb|FQS| happens very late in the run and does not take the full benefit of the superior expressivity of \verb|FQS|. Also, the switching threshold $E_t$ plays a crucial role in the performance of the cost-average-based hybrid, since with larger systems, the ground state might be decreasing as the number of qubits grows, as in the Heisenberg model case. 

Next, we compared the proposed hybrid algorithms to each other with the same data as presented in the bottom sub-figure in Fig.~\ref{hybrid_scalability_results_early_stop}. Here, the early stopping is a better choice overall regardless of the system size. If we are uncertain what threshold $E_t$ to set for early stopping, it does not have a significant impact on the outcome, and should yield better results than any optimizer used alone. For the cost-average-based hybrid, we are more concerned with the window length $w$ to compute the average of the last evaluations of the cost function. From Fig.~\ref{averages_differences_heisenberg}, we get information on how the difference between the new cost value is compared to the average computed for the last $w$ cost values. The difference scales according to the length of the window, so we can keep the switching threshold $E_t$ fixed and only focus on finding the best $w$. In our case, for the Heisenberg model, we prefer a small window length regardless of the system size in order to obtain the best possible results for the cost-average hybrid algorithm.

In Fig.~\ref{ALL_HYBRIDS_scalability_results}, we compare the scalability of \verb|RotoFQS| to gate-wise and iteration hybrids. We set $w=10$ for the cost-average hybrid with the threshold $E_t=0.01$. The early stopping hybrid patience was set to $P=5$ with the threshold $E_t=0.1$. For gate-wise hybrids, we used $p=0.25$ and $p=0.5$, and in the iteration hybrid, we set $N=2$. Across the different system sizes, the cost function hybrids \verb|RotoFQS| exhibit much better scalability than gate-wise and iteration hybrids. As the system size grows, the better the cost function hybrids tend to perform compared to the gate-wise and iteration hybrids. In addition, the gate and iteration hybrids require more classical resources due to the frequent change of representation for the gates in PQC. As we performed the simulations for larger systems (13 and 15 qubits), we noticed a huge difference in the simulation time of gate-wise hybrids and cost function hybrids. The gate-wise hybrids required significantly more time to simulate than the other hybrids. 

\subsection{Fidelity Maximization}

In our final experiment, we intend to maximize the fidelity with a randomly sampled $n$-qubit quantum state $\ket{\phi}$. The fidelity of two pure quantum states $\ket{\phi}$ and $\ket{\psi}$ is defined as 
\begin{equation}
    F(\ket{\phi}, \ket{\psi}) = |\langle\phi|\psi \rangle|^2.
\end{equation}
Another metric that can measure fidelity is trace distance $T(\ket{\phi}, \ket{\psi})$, which can be expressed from the fidelity formula of the previous equation as follows
\begin{equation}
    T(\ket{\phi}, \ket{\psi}) = \sqrt{1 - F(\ket{\phi}, \ket{\psi})}.
\end{equation}
Since the main goal is to maximize the fidelity, it is the same as minimizing the trace distance, which we use as a metric in this work. 

To maximize the fidelity of a random quantum state $\ket{\phi}$ and the quantum state $\ket{\psi_{\bm{\theta}}}$ produced by the PQC with parameters $\bm{\theta} = (\theta_1,\ldots,\theta_{Ln})$, we define a projection operator $\mathcal{P} = - \ket{\phi} \bra{\phi}$ from the target state $\ket{\phi}$, which we use as a Hermitian observable in the cost function. The cost function $\expval{M}_{\bm{\theta}}$ that we minimize is expressed with the fidelity or the trace distance as follows
\begin{equation}
    \expval{M}_{\bm{\theta}} = - F(\ket{\phi}, \ket{\psi_{\bm{\theta}}}) =  T(\ket{\phi}, \ket{\psi_{\bm{\theta}}})^2 - 1.
\end{equation}

We examined the performance of \verb|Rotosolve|, \verb|Fraxis|, and \verb|FQS| compared to hybrid methods in which we used window length $w=3,5$ and 10 for the cost-average hybrid with switching thresholds $E_t=0.1$ and $E_t=0.01$. For the early stopping hybrid, we used patience $P=5$ and thresholds $E_t=0.1, 0.05, 0.01$.  In contrast to previous experiments, we used the same number of iterations as in previous sections, but now we performed 50 runs for each optimizer for 4-qubit random state fidelity maximization with 4 layers for the ansatz circuit. At the beginning of every run, we randomly sample a new $n$-qubit target state $\ket{\phi}$. The target state is fully sampled in the following way: First, we sample $2^n$ random complex numbers $z_k = x_k + y_ki$ from the standard normal distribution $\mathcal{N}(0,1)$, where the mean is zero and the standard deviation is set to unity. The coefficients $x_k$ and $y_k$ are sampled separately. After that, we form the quantum state $\ket{\Psi}$ of size $2^n$ and then normalize it to have a unit length $\ket{\phi} = \ket{\Psi} / \norm{\Psi}^2$. Finally, we then create the projection operator $\mathcal{P} = -\ket{\phi} \bra{\phi}$, which we use as the Hermitian observable in the cost function.

\begin{figure}
    \centering
    \includegraphics[width=0.99\linewidth]{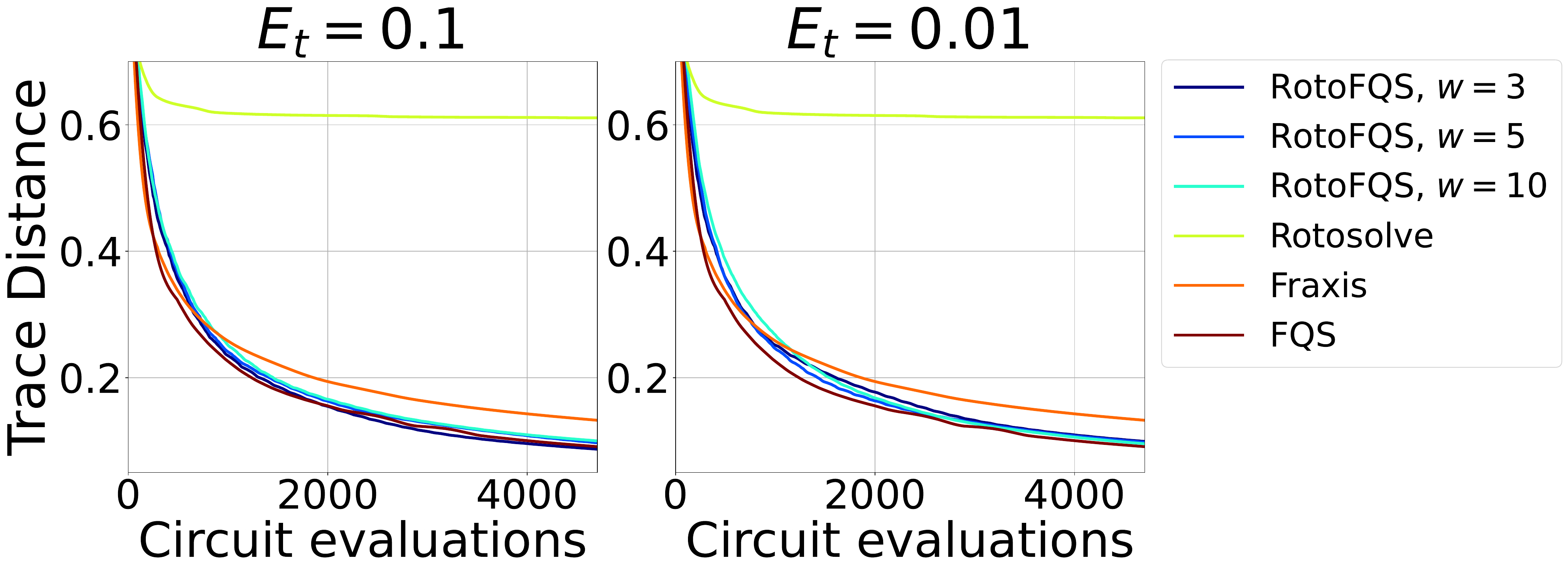}
    \includegraphics[width=0.99\linewidth]{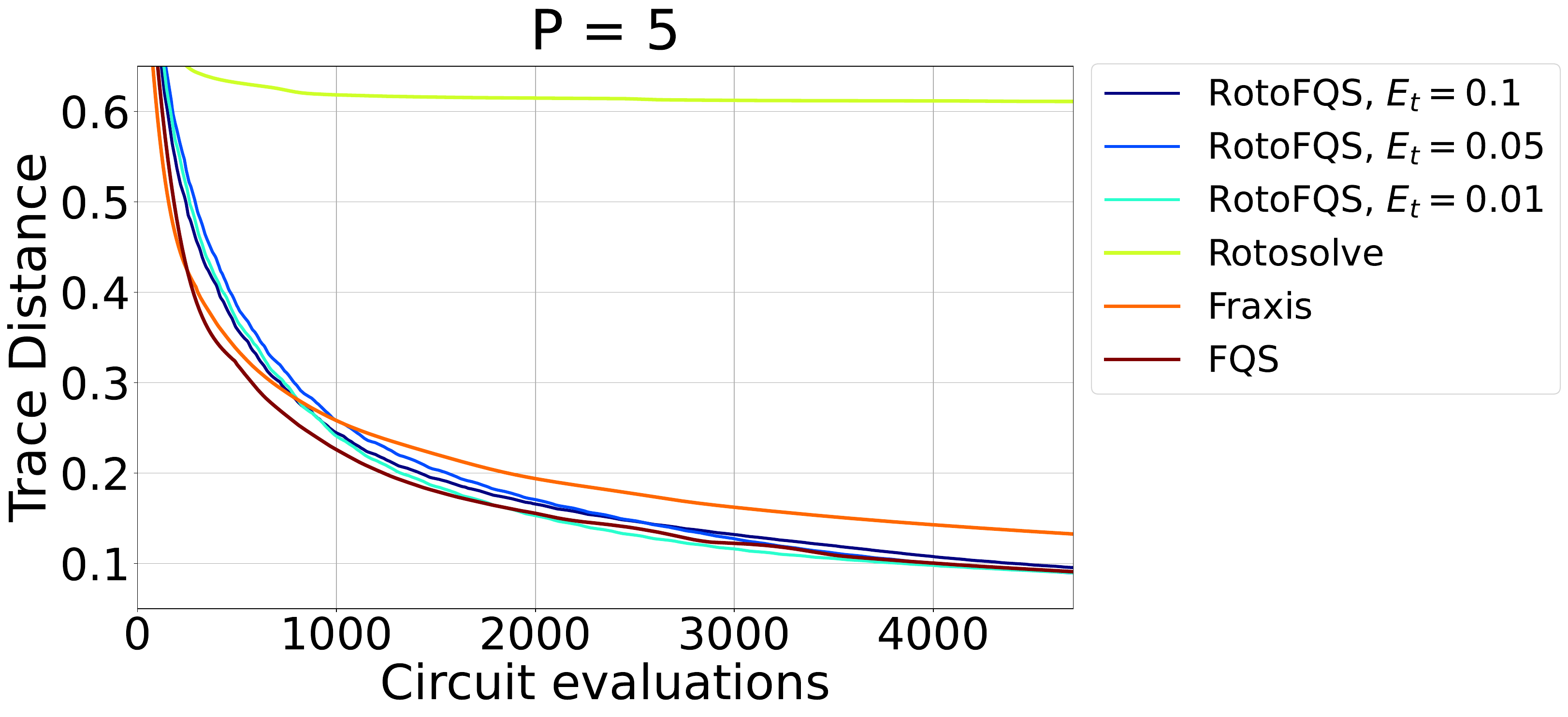}
    \cprotect\caption{Results for fidelity maximization of 4-qubit quantum states with 4 layers for \verb|Rotosolve| (yellow), \verb|Fraxis| (orange), \verb|FQS| (dark red) optimizers compared to \verb|RotoFQS| hybrid. The bottom sub-figure represents results for the early stopping hybrid of \verb|Rotosolve| and \verb|FQS| with patience set to $P=5$ with thresholds $E_t = 0.1, 0.05$ and $E_t = 0.01$. The top row represents the results for the cost-average-based hybrid method of \verb|Rotosolve| and \verb|FQS| with switching thresholds $E_t=0.1$ (left) and $E_t=0.01$ (right). The window lengths were set to $w=3,5,10$. Each line represents the mean of 50 runs.}
    \label{random_state_rotofqs_results}
\end{figure}

We present the results for \verb|RotoFQS| used with early stopping and cost-average hybrids in Fig.~\ref{random_state_rotofqs_results}. Compared to previous experiments with these hybrids in \verb|RotoFQS| we used small window lengths, as the cost function is bounded in the range $[0,1]$ and the random state is challenging to optimize. This leads to a plateau of the optimizer \verb|Rotosolve| relatively quickly compared to other cost functions used in the experiments. The results for the cost-average hybrid of top sub-figures in Fig.~\ref{random_state_rotofqs_results} show that even though using different combinations of window lengths $w$ and thresholds $E_t$, they cannot surpass the \verb|FQS| optimizer due to its superior expressivity. In all sub-figures, the \verb|FQS| has a better convergence speed, but the hybrids still begin to converge toward its performance in all cases, and both hybrids. Other standalone optimizers \verb|Rotosolve| and \verb|Fraxis| are outperformed by the hybrids as well. 

\begin{figure}
    \centering
    \includegraphics[width=0.99\linewidth]{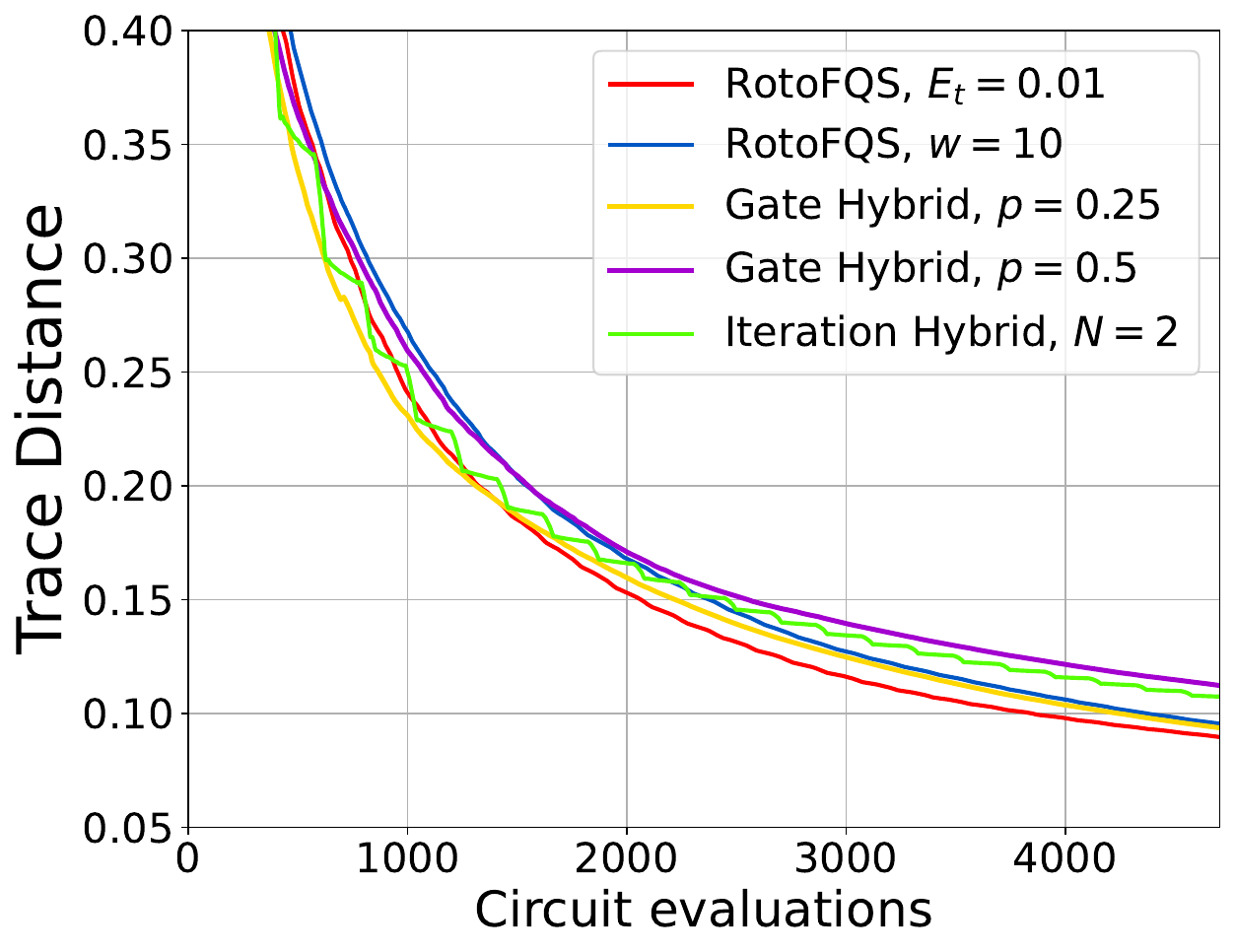}
    \cprotect\caption{A comparison of all hybrid methods: early stopping (red), cost-average (blue), gate-wise with $p=0.25$ (yellow), $p=0.5$ (purple), and iteration hybrid (green) with $N=2$ for fidelity maximization of 4-qubit quantum states with 4 layers. The patience of the early stopping hybrid algorithm was set to $P=5$, and the window length $w=10$ was set for the cost-average hybrid (blue), and in both $E_t=0.01$. Each line represents the mean across the 20 runs.}
    \label{ALL_HYBRIDS_random_state_results}
\end{figure}

In Fig.~\ref{ALL_HYBRIDS_random_state_results}, we examine the convergence between all hybrid algorithms. For the cost function hybrids, we set patience to $P=5$ for early stopping, and a window length of $w=10$ is used for the cost average hybrid. Again, we used a threshold $E_t=0.01$ for both cost function hybrids. Parameters for gate-wise hybrid were set to $p=0.25$ and $p=0.5$, and for iteration hybrid, $N=2$ was used as in the previous experiments. Contrary to the previous experiments with the Heisenberg and Fermi-Hubbard Hamiltonians, all hybrids exhibit nearly the same convergence in terms of fidelity maximization. The gate hybrid with $p=0.5$ performs the worst, followed by the iteration hybrid. Then the cost average hybrid and gate hybrid with $p=0.25$ perform equally well in the end, but the gate hybrid has a better initial convergence. Finally, the early stopping hybrid narrowly has the best mean across 50 individual runs, but that is negligible if we look at it on the scale of Fig.~\ref{random_state_rotofqs_results}.

\section{Conclusions} \label{conclusion_section}

In this work, we proposed two new hybrid algorithms based on the measured cost function values, one based on the early stopping method from classical ML, and the other based on the cost function averages across different window sizes.  This further enhances and expands the hybrid algorithms compared to previous work in Ref.~\cite{pankkonen2025improvingvariationalquantumcircuit}, where hybrid methods were based on iterations or a probabilistic way to choose which optimizer to use at the gate level. The hybrid algorithms in this work consisted of \verb|Rotosolve| and \verb|FQS| outperform the single-qubit optimizers \verb|Rotosolve|, \verb|Fraxis|, and \verb|FQS| used independently in most experiments. In addition, the new hybrid methods based on the cost function values are significantly better than the probabilistic gate-wise hybrid and iteration hybrid method from the previous work. Our proposed hybrid algorithms are based on the measured values of the cost function and are accompanied by a threshold $E_t$ and additional logic with patience for early stopping and window length for cost averages when to switch from a less expressive optimizer $\mathcal{A}$ to a more expressive optimizer $\mathcal{B}$. 

The numerical experiments of the 10-qubit Heisenberg model show that with both an ideal and a noisy quantum device, the early stopping hybrid has the best performance compared to other optimizers. When a noisy device is used, the results are even better compared to the results of an ideal device. The cost-average-based hybrid performed the best with a low window length across the different algorithm switching thresholds $E_t$. Additionally, the cost function hybrids outperformed gate and iteration hybrids in terms of convergence speed and finding better minima. 

Then we examined 6-qubit Fermi-Hubbard on a $1\times 3$ lattice with 5 layers across all optimizers. Again, the hybrids demonstrate the robustness with different noisy profiles of the quantum devices of 2048, 4096, and 8192 shots. Hybrids composed of \verb|Rotosolve| and \verb|FQS| obtained the best results with the highest amount of noise produced by the device. That is, with the lowest shot count of 2048, we get the best results that are compatible with the current NISQ era devices. With fewer shots, we can save the computational resources of current quantum computers and still be able to get relatively good results with hybrid algorithms. This is also the case when using the state-vector simulator for the ideal quantum device. Again, we find the best results with a short window length for the cost-average hybrid and a high switching threshold $E_t$ for the early stopping hybrid algorithm. The comparison of all hybrids resulted in the favor of cost function hybrids, but the gate and iteration hybrids had a better initial convergence this time.

We also tested the scalability of our proposed hybrid algorithms with varying system sizes for the Heisenberg model. We specifically focused on the \verb|RotoFQS| hybrid consisting of \verb|Rotosolve| and \verb|FQS| optimizers. The hybrid algorithms demonstrate robust scalability across the system sizes in terms of the number of qubits and layers used in the circuit. As the number of qubits grows, the performance of the hybrid algorithms gets better compared to the standalone optimizers. We noticed that for the early stopping, when keeping patience $P$ fixed, the switching threshold $E_t$ does not have a significant impact on the overall performance. However, while keeping $E_t$ fixed for the hybrid based on cost-average and varying the window length, it does have a noticeable impact. As we grow $w$, the performance starts to resemble more and more \verb|Rotosolve|. That is, the cost-average-hybrid across different cost functions and circuit sizes benefits the most when $w$ is set to a relatively small value. With larger system sizes, there was a significant difference between the cost function and previously proposed hybrids. Especially with a 15-qubit system with 15 layers, the new hybrids have better scalability. In addition, we noticed a big downside of the gate and iteration hybrids. As the system size and the number of gates in the PQC grow, we need more and more classical computation resources to switch the individual gates from one representation to another. This adds additional cost in time for the optimization tasks. This is crucial when we scale the system sizes from tens of qubits to hundreds of qubits. With that in mind, the cost function-based hybrids proposed in this work alleviate this problem as we switch the gate representation from one to another only once in the optimization process.

Finally, we examined the hybrid algorithms' performance on fidelity maximization for a 4-qubit system. For \verb|RotoFQS|, the hybrids were not able to have a better performance compared to \verb|FQS|, so in fidelity maximization, it is better to use \verb|FQS| alone without hybrid implementation. When compared to gate and iteration hybrids, there was no significant difference in which hybrid was used in the fidelity maximization, as they all performed equally well. 

To conclude our work, we emphasize the scalability and robustness of the proposed hybrid methods for different cost functions compared to the hybrid methods proposed in previous work~\cite{pankkonen2025improvingvariationalquantumcircuit}. Our proposed hybrid methods can be extended to have multiple optimizers. That is, when the optimization begins with an optimizer $\mathcal{A}$, and after switching to an optimizer $\mathcal{B}$, we could further switch the optimizer from $\mathcal{B}$ to an even more expressive optimizer $\mathcal{C}$ when a given criterion is met. Here, optimizers are used from least expressive to most expressive or cost-heavy, that is, $\mathcal{A} < \mathcal{B} < \mathcal{C}$, where $\mathcal{A}$ is the least expressive optimizer and $\mathcal{C}$ is the most expressive optimizer. This could further improve the speed of convergence towards the global minima of the cost function used in the optimization and would be interesting to study in detail.

\begin{acknowledgments}
We acknowledge funding by Business Finland for the project 8726/31/2022 CICAQU. J.V.P. received funding from InstituteQ's doctoral school. 
\end{acknowledgments}

\section*{Data Availability}

The code and data to generate the figures in this work are available upon reasonable request.  

\onecolumngrid
\appendix

\section{Average Cost Function Values for Optimizers}\label{appendix_window_lengths}

\begin{figure*}
    \centering
    \includegraphics[width=0.99\linewidth]{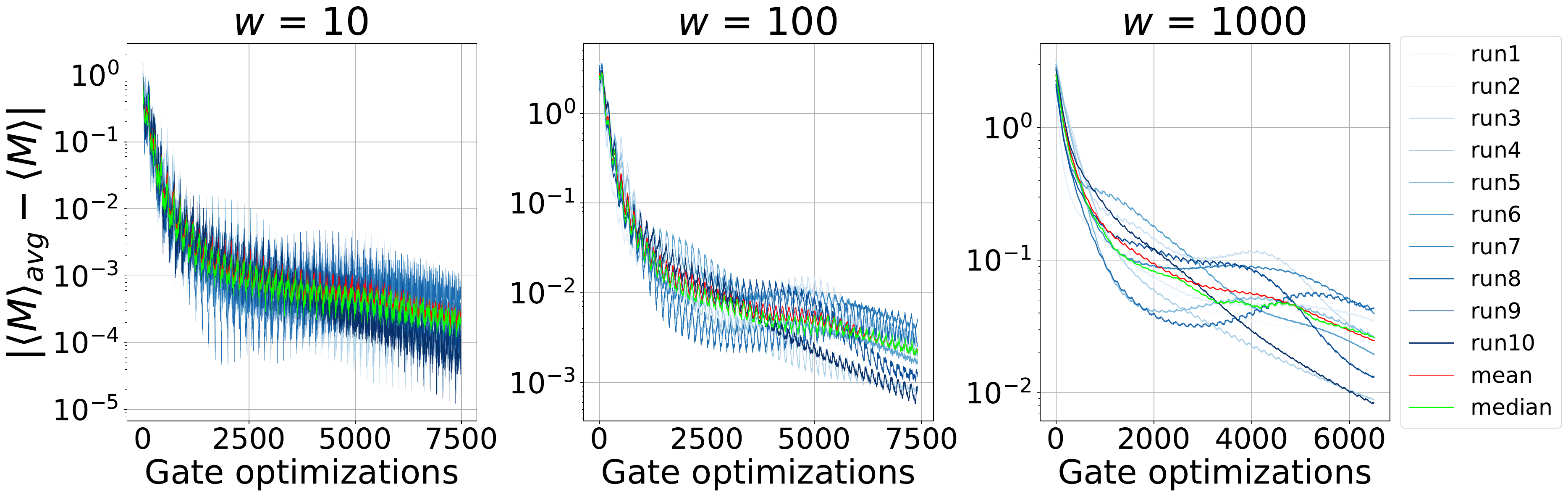}
    \includegraphics[width=0.99\linewidth]{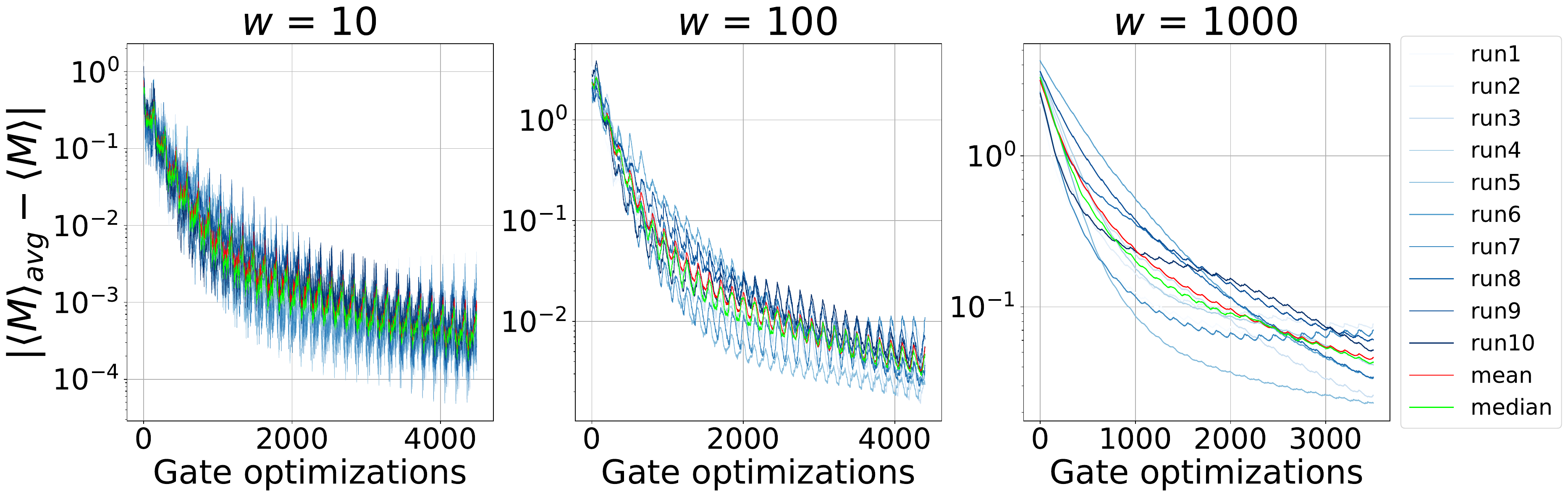}
    \cprotect\caption{Differences between average cost function $\expval{M}_{avg}$ values across gate optimizations compared to the new cost function value $\expval{M}$ on a logarithmic scale. Results are from a 10-qubit Heisenberg model with 15 layers. \verb|Fraxis| (top row) and \verb|FQS| (bottom row) optimizers were used. Each run is plotted in different shades of blue, the red line is the mean, and the green line represents the median. The averages $\expval{M}_{avg}$ are computed from the latest $w$ gate optimizations. On the left, $w=10$ was used, $w=100$ on the middle plot, and $w=1000$ on the right plot.}
    \label{averages_differences_heisenberg_fraxis_fqs}
\end{figure*}

We present additional results for cost function averages compared to the newest cost function value across the different average window lengths $w$ and optimizers, similar to Fig.~\ref{averages_differences_heisenberg} in Sec.~\ref{results_section}. In Fig.~\ref{averages_differences_heisenberg_fraxis_fqs}, we present similar results for \verb|Fraxis| and \verb|FQS| with a 10-qubit Heisenberg model and 15 layers. The data is taken from the first 10 runs of the Fig.~\ref{hybrid_heisenberg_results} for both optimizers with an ideal quantum device simulation. Compared to \verb|Rotosolve| in Fig.~\ref{averages_differences_heisenberg}, \verb|Fraxis| and \verb|FQS| they exhibit a similar behavior as the function of gate optimizations performed on the circuit across the window lengths of the taken average. When the $w$ is assigned smaller values, we obtain smaller and smaller differences faster during the optimization, and larger differences with larger $w$. Regardless of the optimizer, the scale of differences $|\expval{M}_{avg} - \expval{M}|$ remains the same across different values of $w$ when they are compared to each other on the number of gate optimizations done.

\begin{figure*}
    \centering
    \includegraphics[width=0.49\linewidth]{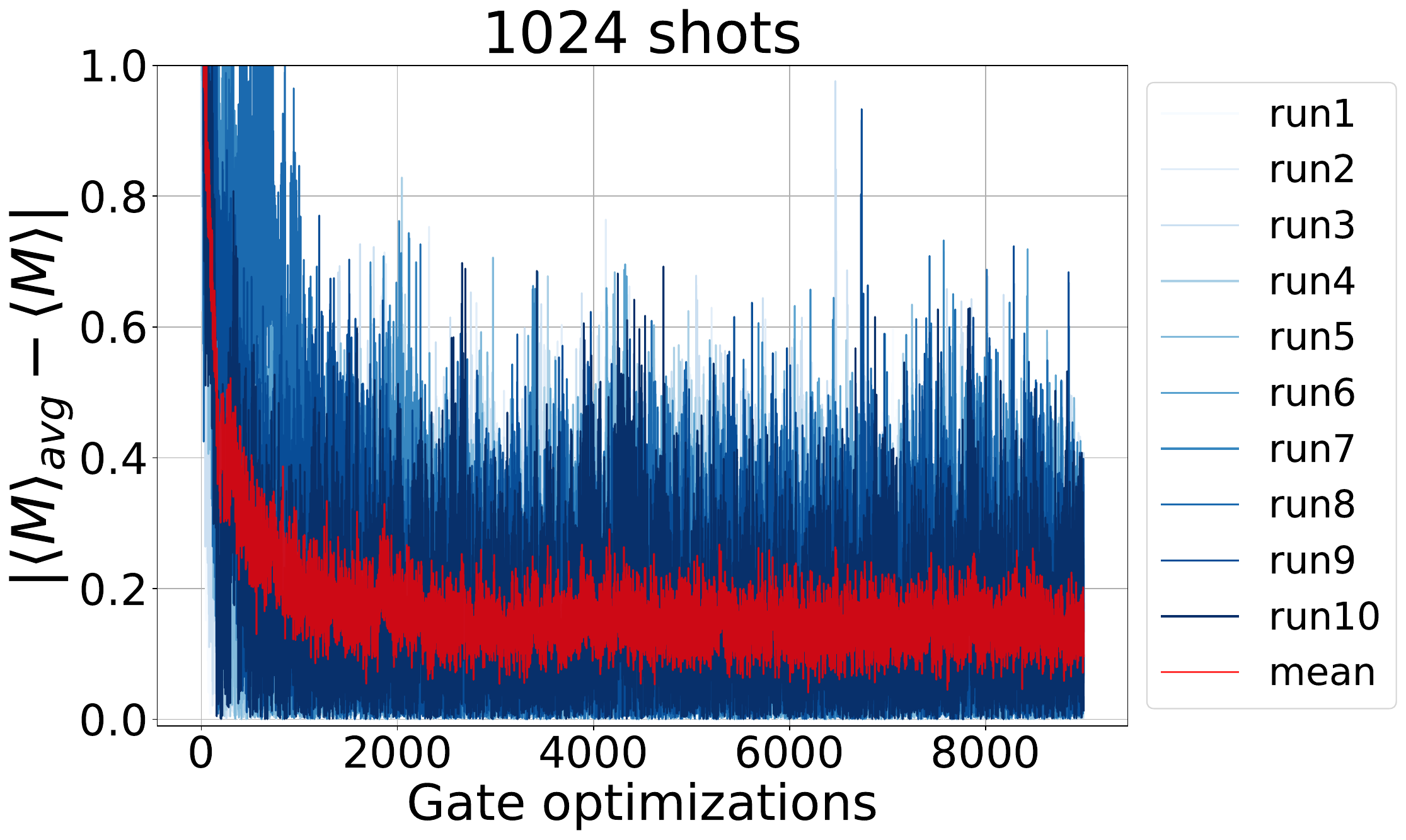}
    \includegraphics[width=0.49\linewidth]{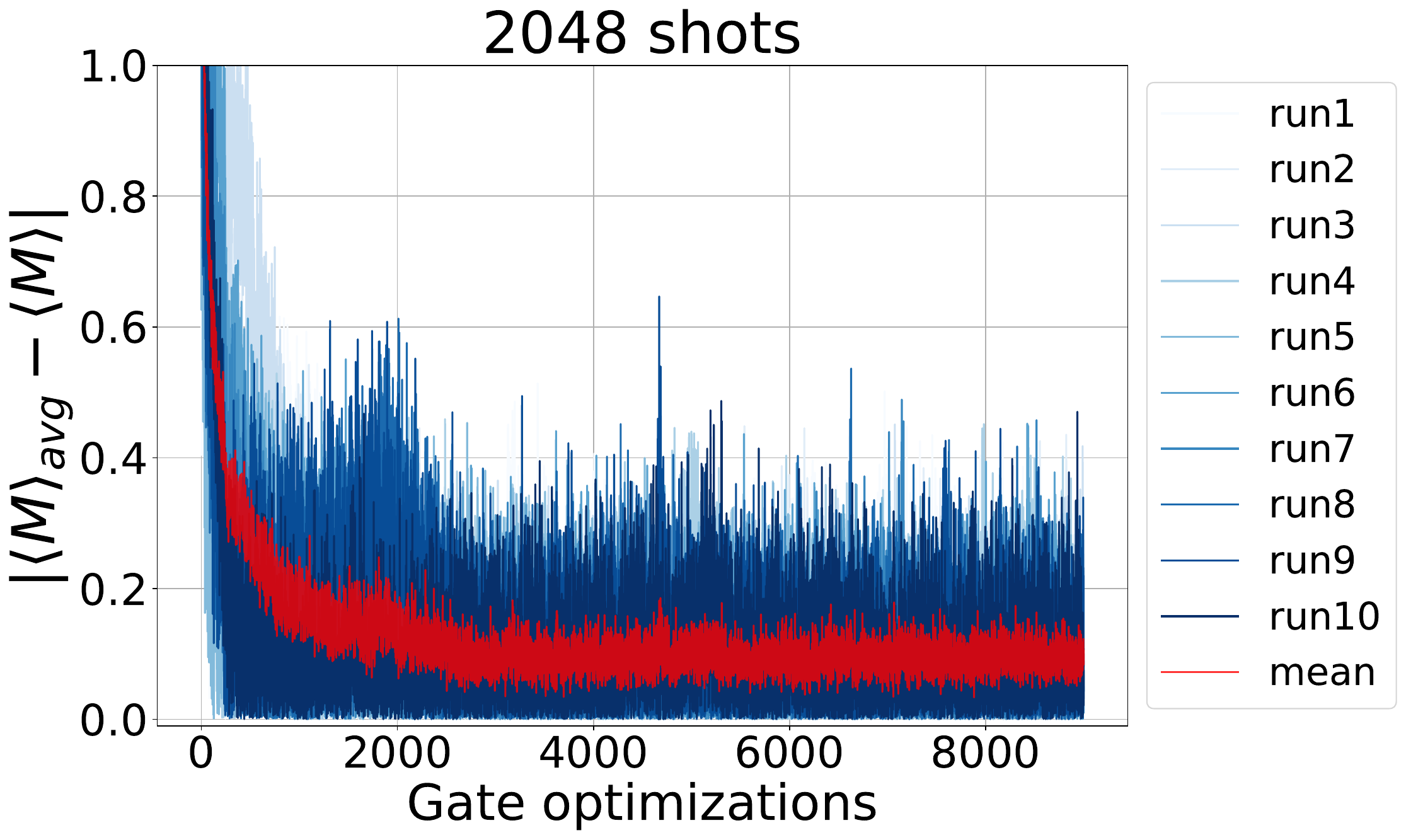}
    \includegraphics[width=0.49\linewidth]{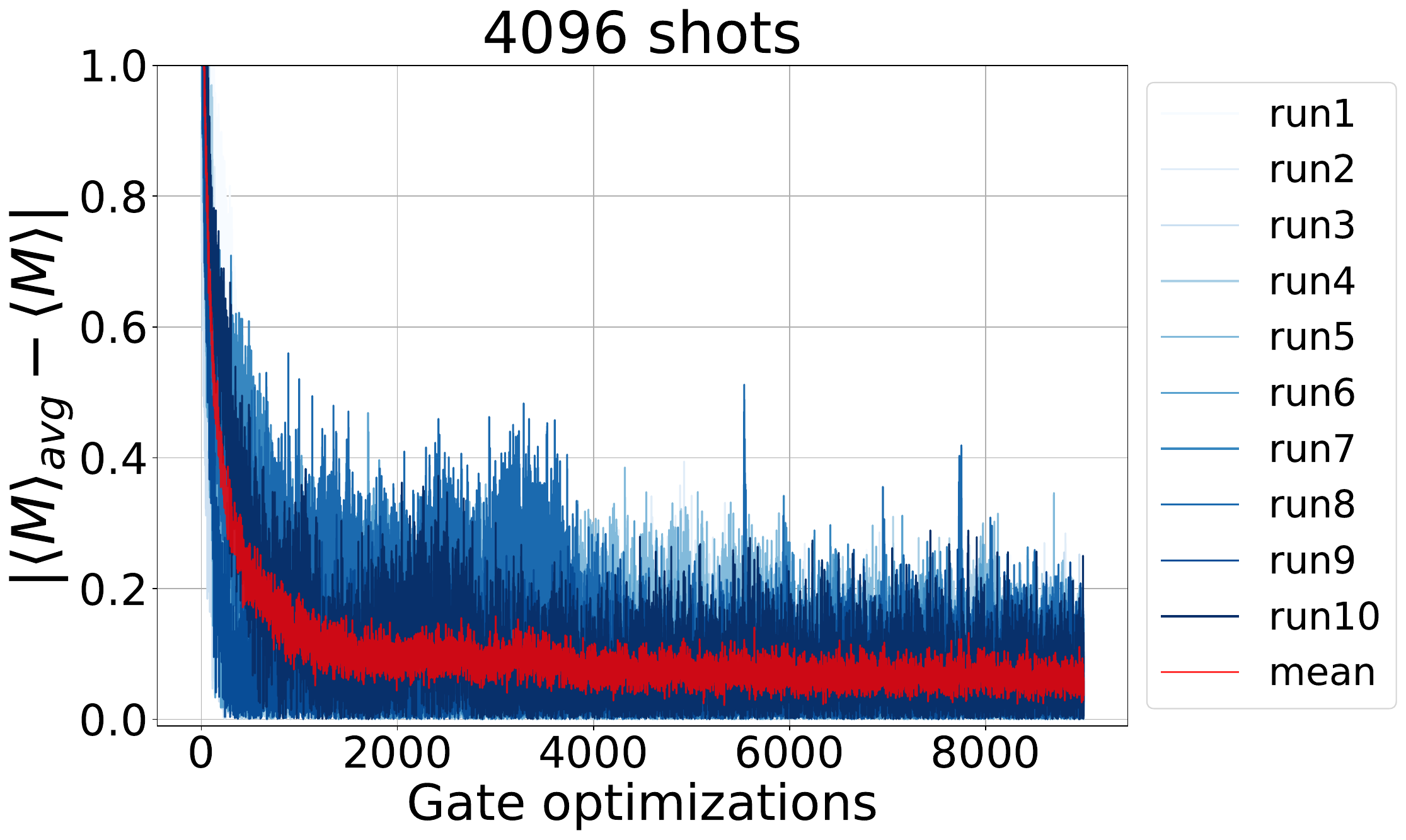}
    \includegraphics[width=0.49\linewidth]{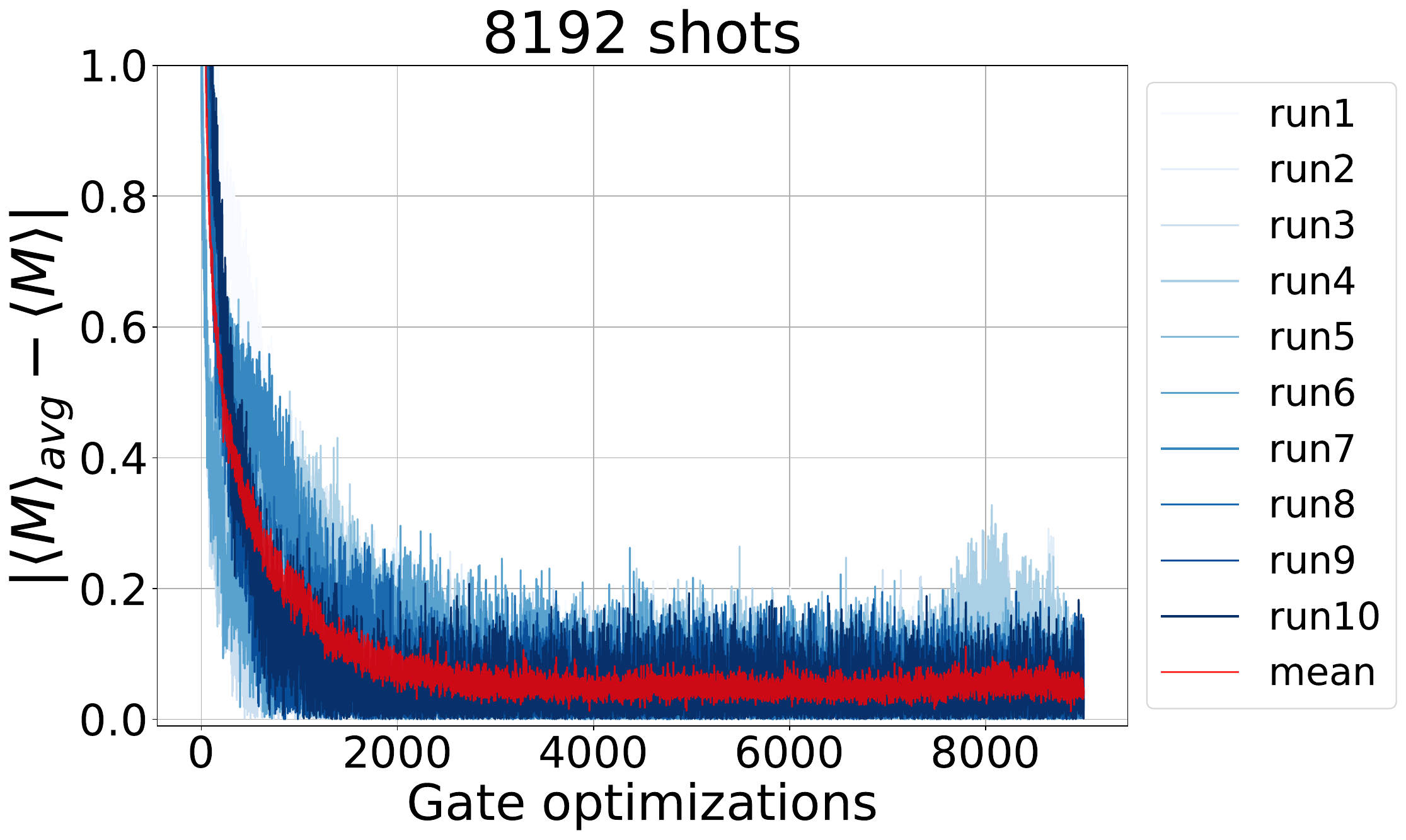}
    \cprotect\caption{Differences between average cost function $\expval{M}_{avg}$ values across gate optimizations compared to the new cost function value $\expval{M}$ on a linear scale. Results are from a 10-qubit Heisenberg model with 10 layers. Each sub-figure represents runs performed with the \verb|Rotosolve| optimizer, using shot counts of 1024 (top left), 2048 (top right), 4096 (bottom left), and 8192 (bottom right) to approximate each Hamiltonian term. Each run is plotted in different shades of blue, and the red line is the mean of 10 runs. The averages $\expval{M}_{avg}$ are computed from the latest $w$ gate optimizations, where $w=1000$.}
    \label{averages_differences_noisy}
\end{figure*}

Additionally, we present the results for the averages of the cost function with a window length of $w=1000$ on noisy devices. We used the 10-qubit Heisenberg model with 10 layers with the \verb|Rotosolve| optimizer. The shots for the noisy device were set to 1024, 2048, 4096, and 8192 shots. The results for the cost function averages are fully displayed in Fig.~\ref{averages_differences_noisy}. With 1024 shots, the difference to the mean $|\expval{M}_{avg} - \expval{M}|$ constantly spikes between 0.5 and 0.6, where the mean across 10 runs for $|\expval{M}_{avg} - \expval{M}|$ plateaus near 0.2. When we set the device to 2048 shots, the individual runs spike maximum at 0.4 and mean 0.1. By further increasing the shots used by the device, i.e., making more accurate measurements, the mean of the 10 runs for $|\expval{M}_{avg} - \expval{M}|$ also decreases and becomes more concentrated. That is, with a more accurate device or more shots, we can make the best use of the cost-average-based hybrid and get more certain about the best switching threshold $E_t$ to a more expressive optimizer.

\bibliography{refs}

\end{document}